\documentclass[12pt,preprint]{aastex}

\slugcomment{}

\shorttitle{Recombination Lines in GEHRs}
\shortauthors{Esteban et al.}

\received{2002 May 15}
\begin{document}

\title{Optical Recombination Lines of Heavy-elements\\
     in Giant Extragalactic H~II Regions\footnotemark{}}

\author{C\'esar Esteban}
\affil{Instituto de Astrof\'\i sica de Canarias, E-38200 La Laguna, Tenerife, 
Spain}
\email{cel@ll.iac.es}

\author{Manuel Peimbert and Silvia Torres-Peimbert}
\affil{Instituto de Astronom\'\i a, UNAM, Apdo. Postal 70-264, M\'exico DF, 
Mexico}
\email{peimbert@astroscu.unam.mx, silvia@astroscu.unam.mx}

\and

\author{M\'onica Rodr\'\i guez}
\affil{INAOE, Apartado Postal 51, Puebla, Pue, Mexico}
\email{mrodri@inaoep.mx}

\begin{abstract}

We present high resolution observations of the giant extragalactic \ion{H}{2} regions NGC 604, NGC 2363, NGC 5461 and
NGC 5471, based on observations taken with the ISIS spectrograph on
the William Herschel Telescope. We have detected -by the first time- C II and O II recombination 
lines in these objects. We find that recombination lines give larger C$^{++}$ 
and O$^{++}$ abundances than collisionallly excited lines, suggesting 
that temperature variations can be present in the objects. We detect  [\ion{Fe}{4}] 
lines in NGC 2363 and NGC 5471, the most confident detection of optical 
lines of 
this kind in \ion{H}{2} regions. Considering the temperature
structure we derive their H, He, C, N, O, Ne, S, Ar, and Fe abundances.
{From} the recombination lines of NGC 5461 and NGC 5471 we determine the presence
of C/H and O/H gradients in M101.
We calculate the  $\Delta Y$/$\Delta O$ and $\Delta Y$/$\Delta Z$ values
considering the presence of temperature variations and under the assumption
of constant temperature. We obtain a better agreement with models of
galactic chemical evolution by considering the presence of temperature 
variations than by assuming that the temperature is constant in these nebulae.

\end{abstract}


\section{Introduction}

The analysis of the spectra of \ion{H}{2} regions allows to determine the 
abundances of He, C, N, O, Ne, S, Ar and Fe in the ionized phase of the 
interstellar medium. This is useful to trace the chemical evolution of 
the interstellar gas, to 
compute the radial abundance gradients in spiral galaxies and even to estimate 
the primordial helium abundance. Due to the surface brightness of distant
extragalactic \ion{H}{2} regions it is possible to measure 
their line intensities with reasonable accuracy. Therefore, it is essential and
feasible to have confident 
determinations of their chemical composition. 
The possibility to obtain deep spectra of \ion{H}{2} regions with large 
telescopes allows us to detect and measure important faint emission lines. 
Among these, recombination lines (hereafter RLs) of 
heavy element ions are of special interest. The brightest RLs of heavy element ions 
in the optical domain are \ion{C}{2} 4267 \mbox{\AA}\ and those of multiplet 1 
of \ion{O}{2} around 4650 \mbox{\AA}. These are in fact very faint lines that 
have an intensity of the order of 0.001$\times$ $I(H\beta)$. 
These lines can give us a more complete view of the physics and chemical content 
of nebulae and can be used to test if the standard methods for deriving chemical 
abundances --based on the intensity of bright collisionally excited 
lines (hereafter CELs)-- are valid. 
 
The ionic abundances of elements heavier than He are usually derived from 
the intensity of CELs, which depend exponentially on 
the electron temperature ($T_e$) of the nebular gas. This fact makes 
necessary to have a very precise determination of $T_e$ to obtain reliable 
ionic abundances. \citet{pei67} found that in the presence 
of inhomogeneities or stratification in the spatial distribution of 
$T_e$ (the so-called 
temperature fluctuations, defined by the mean square temperature variation 
over the observed volume: $t^2$) the ionic abundances obtained from 
the intensity of CELs are systematically underestimated. In comparison, 
ionic abundances determined from RLs are almost independent on $T_e$ 
and are not sensible to the effects of possible temperature structure 
inside the nebula. However, the faintness of these lines makes very difficult 
their measurement and even their detection. \citet{est98,est99a,est99b} 
have obtained high resolution observations of 
the Galactic \ion{H}{2} regions Orion nebula, M8, and M17, obtaining 
good measurements of \ion{C}{2} and \ion{O}{2} lines in the three objects. 
These authors have found that ionic abundances derived from those RLs 
are systematically larger than the values obtained from CELs. A similar 
result has been obtained by \citet{tsa03} who 
present measurements of \ion{C}{2}, \ion{N}{2}, and \ion{O}{2} lines for  
Orion nebula, M17, NGC 3576, and three Magellanic Clouds \ion{H}{2} regions (30 Doradus, LMC N11, and 
SMC N66).

The main aim of the observations reported in this paper was to detect and 
measure \ion{C}{2} and \ion{O}{2} lines in bright 
giant extragalactic \ion{H}{2} regions (hereafter GEHRs) of the northern 
hemisphere. These observations will permit to compare the O$^{++}$ 
abundances obtained by both CELs and RLs from the same spectrum as 
well as to derive the C$^{++}$ abundance and compare them with the values 
derived by other authors from space observations of the  UV [\ion{C}{3}] 
1907 $+$ \ion{C}{3}]  1909 \mbox{\AA}\ lines. 

\footnotetext{Based on observations made with William Herschel 
Telescope operated on the island of La Palma by the Isaac Newton 
Group of Telescopes in the 
Spanish Observatorio del Roque de Los Muchachos of the 
Instituto de Astrof\'\i sica de Canarias.}

\section{Observations}

The observations were made on 2001 February 25 and 26 at the 
Observatorio del Roque de los Muchachos (La Palma), using the 4.2 m 
William Herschel Telescope (WHT) with the ISIS spectrograph at 
the Cassegrain focus. Two different CCDs were used at the blue and red 
arms of the spectrograph: an EEV12 CCD with a configuration of 
4096$\times$2048 pixels of 13 $\mu$m in the blue arm and a 
TEK with 1024$\times$1024 of 24 $\mu$m in the red arm. 
The spatial scale was 0.19 arcsec pixel$^{-1}$ and 0.36 arcsec 
pixel$^{-1}$ for the blue and red arms respectively. 
The maximum unvignetted slit-length usable with ISIS is 3.7$'$ and 
the width was 1$''$.  The dichroic used to separate the blue and red beams
was set at 5400 \mbox{\AA}.  Two gratings were used, a 1200 g mm$^{-1}$ one
in the blue and a 316 g mm$^{-1}$ one in the red arm.  These gratings gave
reciprocal dispersions of 17 and 62 \mbox{\AA} \ mm$^{-1}$, and
effective resolutions of 0.73 \mbox{\AA} \ and 2.9 \mbox{\AA} \ for the blue and red
arms, respectively. The blue spectra cover from 4230 \mbox{\AA}\ to 5060 \mbox{\AA}\ 
and the red ones from 5720 \mbox{\AA}\ to 7200 \mbox{\AA}. A summary of the 
observations is presented in Table 1. 

The average seeing was between 0.8$''$ and 2$''$ throughout the 
observations.  The slit
center and position angle (PA) was chosen to 
cover the brightest zone of each \ion{H}{2} region.  For 
NGC~604 we centered our slit on zone D defined by \citet{dod83} that was 
also observed by \citet{dia87}. NGC~5461 is rather compact and we placed 
the slit center on the bright core passing through two fainter 
\ion{H}{2} regions located at 15 and 30 arcsec to the southeast 
\citep[see image shown by][]{gia99}. For NGC~5471 we centered 
the slit passing through the knots labeled as A and C by 
\citet[][see his Figure 1]{ski85}. Finally, for NGC~2363 our 
slit position covered knots A 
and B defined by \citet[][see their Figure 2]{gon94}. 
Several 30 min exposures were combined to produce the final blue and red
spectra of the objects. As we can see in Table 1, the total exposure time 
for the objects was very large (between 2.5 and 7 hours), this 
indicates that we are at the limit of the possibilities of a 4m-class 
telescope. Comparison lamp exposures (CuAr + CuNe) were taken 
between consecutive spectra of the same object. The absolute flux 
calibration was achieved by observations of the standard stars Feige 15, 
Feige 25, Feige 56 \citep{sto77}, Feige 34, G191 B2B \citep{mas88}, and 
BD+33 2642 \citep{oke90}. The correction for atmospheric extinction 
was performed using an average curve for the continuous atmospheric 
extinction at Roque de los Muchachos Observatory. 

The data were reduced using the standard $IRAF$\footnotemark{} 
LONGSLIT reduction package to perform bias corrections, flatfielding,
cosmic ray rejection and flux calibration.  

\footnotetext{IRAF is distributed by NOAO, which is operated by AURA, under
cooperative agreement with NSF.}

\section{Line Intensities and Physical Conditions}

Line intensities were measured integrating all the 
flux in the line between two given limits and over a local continuum 
estimated by eye. In the few cases of line-blending, the line flux of each 
individual line was derived from a multiple Gaussian profile 
fit procedure. All these measurements were made with the SPLOT routine of 
the $IRAF$ package. 

All the line intensities of a given spectrum have been normalized to a 
particular bright recombination line present in each wavelength interval. 
For the blue spectra the reference line was H$\beta$. For NGC~604, NGC~5461 
and NGC~2363 it was necessary to take shorter exposures in the red arm to 
avoid problems of saturation of H$\alpha$.  
In these three objects we normalized the line intensities measured in the 
longer exposure spectra to \ion{He}{1} $\lambda$ 5876 \mbox{\AA}. 
Finally, the intensity ratios were re-scaled to the 
$I$(\ion{He}{1} 5876)/$I$(H$\alpha$) ratio obtained from the shorter 
exposure red spectra. 

The reddening coefficient, $C$(H$\beta$), was determined 
by fitting iteratively the observed $I$(H$\beta$)/$I$(H$\gamma$) ratio to 
the theoretical one computed by \citet{sto95} for the 
physical conditions determined for each object (see below) and assuming 
the extinction law of \citet{sea79}.  The Balmer line intensities
were not corrected for the presence of underlying stellar absorption lines
due to their large equivalent widths in emission. In Table 2 we present the
$EW$(H$\beta$) values for all  the objects. Alternatively, the $\lambda \lambda$
4388, 4471, and 4922 \ion{He}{1} lines present in Table 2 were corrected for 
underlying absorption considering the starburst models of \citet{gon99}
and considering that the observed continuum is due to a stellar component
plus a nebular component. 
The observed intensities of these helium lines (corrected for reddening but not for
underlying absorption) were:  0.50, 4.14, and 1.07 for NGC~604;  
0.41, 4.11, and 1.09 for NGC~5461; 0.41, 3.89, and 0.89 for NGC~5471; and
0.46, 3.93, and 0.98 for NGC~2363. The line intensities of the blue 
arm spectra were corrected for reddening with respect to H$\beta$ and the red arm 
ones with respect to H$\alpha$. Finally, the resulting reddening corrected 
line ratios of the red arm spectra were re-scaled to $I$(H$\beta$) simply 
multiplying by the theoretical $I$(H$\alpha$)/$I$(H$\beta$) corresponding to 
the physical conditions derived for each nebula. 

The values of $C$(H$\beta$) obtained are consistent with those obtained by 
other authors for NGC~604 \citep[0.30:][]{dia87} and NGC~5471 
(0.02: Garnett et al. 1999; 0.18: Luridiana et al. 2002), slightly lower in the 
case of NGC~5461 (0.56: Garnett et al. 1999; 0.45: Luridiana et al. 2002), 
and clearly higher in the case of NGC~2363
(0.25: Peimbert et al. 1986; 0.20: Gonz\'alez-Delgado et al. 1994). 
The reason of the large difference found in the case of 
NGC~2363 
is unknown but seems to be a real feature and not an error in the flux 
calibration nor an effect of atmospheric refraction.  
The final spectrum used for NGC~2363 is a combination of 
several spectra taken in two consecutive nights, calibrated independently 
with different standard stars and 
giving a similar value of $C$(H$\beta$) (0.74 for February 25th and 0.66 for 
February 26th). In any case, the effect on the derived intensity ratios 
is not large because the emission lines of the blue arm spectra have 
been corrected for reddening with respect to H$\beta$ and those of the 
red arm with respect to H$\alpha$. Therefore, the  wavelength baseline 
between the lines is not large enough to produce strong intensity 
uncertainties.  By comparing the intensities obtained 
from our  $C$(H$\beta$) value with those obtained from the $C$(H$\beta$) 
value by \citet{gon94} the differences in the line intensities reach a 
maximum of 15\% for the C II 4267 \mbox{\AA}  ,
[\ion{N}{2}] 5755 \mbox{\AA} , and \ion{He}{1} 5876 \mbox{\AA}\ lines and are considerably 
smaller for the other lines.
 
The final list of observed wavelengths (referred to the heliocentric reference frame) and line intensities (including their 
uncertainties) relative to H$\beta$ for all the objects is presented in Table 2. 
Colons indicate uncertainties of the order of or larger than 40 per cent. 
For a given emission line, the observed wavelength is determined by 
the centroid of a Gaussian fit of its line profile and it is included in 
columns 5, 7 , 9, and 11 of Table 2. 

Figures 1 to 4 show the spectra of the four GEHRs observed. Enlargement 
of the spectral zones containing \ion{C}{2}  4267 \mbox{\AA}\ line and the 
lines of multiplet 1 of \ion{O}{2} (around 4650 \mbox{\AA}) are included in the 
figures. It is evident that the \ion{C}{2}  4267 \mbox{\AA}\ line is well measured in 
the spectra of NGC~604 and NGC~5461 and barely detected in 
NGC~2363 and NGC~5471. \citet{gra76} showed that recombination 
dominates the excitation of \ion{C}{2} 4267 \mbox{\AA}\ by an order of magnitude. 
This line comes from a transition involving terms with large $l$ 
values (3$d^2D$-4$f^2F^0$), 
levels that cannot be excited by permitted resonance transitions from the 
ground term (2$p^2D^0$). 
We have detected several lines of different multiplets of \ion{O}{2} 
in the spectra of the GEHRs observed. In most of the cases the lines are at the 
detection level, only those belonging to multiplet 1 --the brightest one-- 
have enough signal-to-noise ratio to be reasonably well measured 
(see Figures 1 to 4). Several lines of multiplet 1 of \ion{O}{2} 
are measured in the objects except in the case of NGC~5471,  
where only one line is detected. It is worth noting that LS-coupling predicts that \ion{O}{2}  4649 \mbox{\AA}\ is the brightest line of multiplet 1, and this is what it is observed 
in NGC~5461, NGC~2363, and also in the Orion nebula \citep{est98}. However, 
in the case of NGC~604 that line is too weak with respect the other lines of the multiplet. 
This anomalous trend has been reported in the spectra of the two slit positions observed 
in M17 by \citet{est99b} as well as in the three Magellanic Cloud  \ion{H}{2}  regions
observed by \citet{tsa03}, but to a lesser extent, in some planetary nebulae \citep{liu03}. 
\citet{liu03} pointed out that this might be caused by the under-population of the ground $^3P_2$ fine-structure level of the recombining O$^{++}$ relative to its statistical 
equilibrium value. 
The excitation mechanism of the 
\ion{O}{2} spectrum was also investigated by \citet{gra76}. The presence of 
many emission lines 
of different multiplets of \ion{O}{2} in the spectra of \ion{H}{2} regions cannot be 
explained by resonance fluorescence, 
in fact some of those observed in NGC~604 come from 4f-3d transitions 
(multiplets 66 and 67) these lines cannot be excited by fluorescence from 
the 2$p^3$$^4S^0$ ground level. Therefore, recombination is the 
dominant excitation mechanism of the observed \ion{O}{2} lines in 
NGC~604, but very probably for \ion{H}{2} regions in general. 

We have observed several lines of \ion{Si}{2} in all the objects. These lines 
belong to multiplets 2 and 4, which correspond to doublets. \citet{gra76} 
has found that starlight excitation of the 5$s^2S$ and 4$d^2D$ terms 
(levels that feed the observed transitions) dominates over recombination 
in producing the observed intensities in \ion{H}{2} regions. Other 
interesting lines observed are those of \ion{Mg}{1}], the strongest 
semiforbidden optical lines in nebular spectra, corresponding to 
$^1S_0-^3P_{1,2}$ transitions. These lines have been observed in bright 
planetary nebulae (PNe) and are always weak, with intensity in the range 
0.05-1.00 per cent of the H$\beta$ intensity 
\citep[see][and references therein]{all88}. \citet{bal00} measured 
the \ion{Mg}{1}] 4571.1 \mbox{\AA}\ line in the Orion nebula (intensity about 0.009 
percent of H$\beta$), as far as we know, the first detection of this line 
reported in an \ion{H}{2} region. \citet{cle87} and 
\cite{all88} have found that the weakness of this line in PNe is due to the 
Mg$^0$+H$^+$ charge-transfer reaction, in the absence of which the 
intensity would be higher. This conclusion can be also applied to the spectra 
of \ion{H}{2} regions.  

In Figures 1 and 2 it can be seen that NGC~604 and NGC~5461 show broad emission 
features around $\lambda$ 4650 \mbox{\AA}\ and $\lambda$ 5800 \mbox{\AA}, 
which are produced by Wolf-Rayet (WR) stars. The bump in the blue spectra 
is produced by the blend of  \ion{He}{2} 4686 \mbox{\AA}, \ion{N}{3} 4640 \mbox{\AA}, and 
perhaps \ion{N}{5} 4620 \mbox{\AA}, characteristic of WN stars. On the other 
hand, the detection 
of the \ion{C}{4} 5808 \mbox{\AA}\ broad emission feature indicates the 
presence of WC stars which can also contribute to a fraction of the 
emission around $\lambda$ 4650 \mbox{\AA}. 
The detection of WR stars in NGC~604 was independently reported by 
\citet{dod81} and \citet{con81}. In their spectroscopical study of 
NGC~604 \citet{dod83} detect WC emission at their zone D, which 
encompasses the brightest part 
of our slit position in this object. \citet{dri93} obtain $HST$ 
images of the ionizing cluster of NGC~604 finding three individual WR stars 
(their WR1, WR2a, and WR2b) in the zone in common with our extracted 
1-D spectrum. In the case of NGC~5461, the presence of the blue WR bump 
around 4650 \mbox{\AA}\ was first detected by \citet{ray82} and later confirmed by 
\citet{dod83}. Our deep spectrum of NGC~5461 has revealed also a faint 
bump corresponding to  \ion{C}{4} 5808 \mbox{\AA}, this indicates -for the 
first time- the presence of WC stars in this region. 

There are several references about the detection of WR stars in NGC~5471, 
\citep[see][and references therein]{sch99}, but we only find faint 
narrow \ion{He}{2} 4686 \mbox{\AA}\ emission in our deep spectrum. 
\citet{izo97} report broad and 
nebular \ion{He}{2} 4686 \mbox{\AA}\ emission in the brightest zone of NGC~2363 
(zone A). However, we only 
find nebular \ion{He}{2} 4686 \mbox{\AA}\ emission in this zone. 

The \ion{O}{2} lines of multiplet 1 are in the problematic spectral zone  
where the WR emission feature is present. This could lead to be suspicious 
of some contamination of stellar emission in the lines identified as 
\ion{O}{2}. However, several reasons lead us to be fairly 
confident that the emission lines identified as belonging to \ion{O}{2} are of 
true nebular nature and not due to stellar emission. Firstly, the detection 
of several lines that can be identified as of \ion{O}{2} (except for 
NGC~5471). Secondly, the observed rest wavelength of all the lines 
are consistent with their identification as \ion{O}{2} ones inside the 
uncertainties in the wavelength calibration. 
Finally, they have a line width similar to that of the rest of the nebular lines. 
In fact the width of the WR bump in NGC~604 and NGC~5461 is clearly 
much broader than the nebular lines. Our relatively high spectral resolution 
permits a proper discrimination between nebular and stellar emission in 
this problematic spectral zone.  

In Table 3 we show the physical conditions 
derived for the four GEHRs from the observed emission line ratios 
and the five-level program for the analysis of emission line nebulae of 
\citet{sha95}. The electron density, $N_e$ has been derived from the ratio 
of the [\ion{S}{2}] 6717, 6731 \mbox{\AA}\ doublet for all the objects 
and also from the [\ion{Ar}{4}] 4711, 4740 \mbox{\AA}\ doublet in the 
case of the higher degree of ionization nebulae (NGC~5471 and NGC~2363). 
The electron temperature, $T_e$ has been derived from both the 
[\ion{O}{3}] and [\ion{N}{2}] emission line ratios. The values of $N_e$ 
and $T_e$ obtained are in good agreement with those reported in 
previous works. In the case of NGC~2363, we report the first determination 
of $T$([\ion{N}{2}]). 

\section{Ionic Abundances from Collisionally Excited Lines}

Ionic abundances of N$^+$, O$^{++}$, S$^+$, S$^{++}$, Ar$^{++}$, 
Ar$^{3+}$, and K$^{3+}$ have been obtained from the intensity of CELs, 
using the five-level atom program of \citet{sha95}, 
and the atomic parameters referenced in it. 
We have measured several  [\ion{Fe}{3}] lines 
in all the \ion{H}{2} regions and detected [\ion{Fe}{4}] lines in NGC~5471 (one line)  and NGC~2363 (four lines). Although  [\ion{Fe}{3}] lines were 
previously reported  in NGC~5461 \citep{tor89} and NGC~5471 
\citep{ski85}, 
the Fe$^{++}$/H$^+$ ratio has been never determined for 
these objects. Following \citet{rod02}, we have used the \ion{Fe}{3} collision strengths of \citet{zha96} and transition probabilities of \citet{qui96} 
to obtain the Fe$^{++}$ 
abundance. No optical [\ion{Fe}{4}] lines have been previously detected 
with confidence in \ion{H}{2} regions. \citet{izo01} report a 
dubious detection of [\ion{Fe}{4}] 4907 \mbox{\AA}\ in a low resolution spectrum of 
the blue compact dwarf galaxy SBS 0335-052 and 
\citet{rub97} 
report the detection of  [\ion{Fe}{4}] 2837 \mbox{\AA}\ in the UV spectrum 
of the Orion nebula. Therefore NGC~2363 and NGC~5471 are -by now- 
the \ion{H}{2} regions where optical [\ion{Fe}{4}] lines have been most 
clearly identified. 
The  Fe$^{3+}$/H$^+$ ratio has been determined considering all 
collisional and downward radiative transitions from the 33 lower levels 
of Fe$^{3+}$. To make these calculations we have used the collision strengths 
of \citet{zha97} and the transition probabilities recommended by 
\citet{fro98} and those of \citet{gar78} for those transitions not included in that  
reference.
We have assumed a two-zone scheme for deriving the ionic 
abundances, adopting $T$([\ion{O}{3}]) for the high-ionization-potential 
ions  O$^{++}$,  S$^{++}$, Ar$^{++}$, Ar$^{3+}$, K$^{3+}$, and Fe$^{3+}$ 
and $T$([\ion{N}{2}]) for the low-ionization-potential ions N$^+$, S$^+$, 
and Fe$^{++}$. The ionic abundances obtained are listed in Table 4. 

\section{Ionic Abundances from Recombination Lines}

We can derive ionic abundances from RLs for 
C$^{++}$, and O$^{++}$ and they are presented in Table 5. 
\ion{C}{2} 4267 \mbox{\AA}\ is a case-insensitive recombination line 
and it is located in 
a spectral zone without problems of line-blending. These important 
features make  \ion{C}{2} 4267 \mbox{\AA}\ a suitable line to derive a proper 
value of the C$^{++}$ abundance. We have used the \ion{C}{2} $\alpha_{eff}$ 
values recently calculated by \citet{dav00} to derive the C$^{++}$ abundance. 
We have assumed $T$([\ion{O}{3}]) as representative of the zone where 
this ion is located. The C$^{++}$/H$^+$ ratio obtained for each 
object is included in Table 5. There are two previous calculations of the 
effective recombination  coefficient for \ion{C}{2}  4267 \mbox{\AA}\ one by 
\citet{peq91} and other by \citet{pen63}. We have compared the C$^{++}$/H$^+$ 
ratios obtained making use of the different CII $\alpha_{eff}$ values 
finding that \citet{dav00} and \citet{peq91} give virtually the same 
abundances (inside $\pm$0.01 dex) whilst the CII $\alpha_{eff}$ obtained  
by \citet{pen63} gives C$^{++}$/H$^+$ ratios about 0.08 dex higher. 

\citet{sto94} has computed the \ion{O}{2} $\alpha_{eff}$, assuming LS 
coupling and different cases (A, B, and C). We have used these coefficients 
to derive the O$^{++}$ abundances listed in Table 5. The physical conditions 
assumed are the same as for the \ion{C}{2} calculations. The abundances 
determined from multiplet 1 are almost independent of the case but 
case B seems to be more appropriate for quartets \citep[see][]{liu95}. 
Taking into account the faintness of the individual lines of \ion{O}{2} 
we have derived the abundances adding the intensity of all the observed 
lines of multiplet 1 and multiplying it by a correction factor which 
introduces the expected contribution of unobserved lines of the multiplet 
assuming LS-coupling \citep[see][]{est98}. The abundance derived by this 
method reduce the observational errors related with the faintness of the 
lines and minimize the effects of possible departures from LS-coupling, as it has been noticed to occur in several objects considering the weakness of \ion{O}{2} 4649 \mbox{\AA}\ line with respect the other ones of the same multiplet (see Sect. 6). 

\section{Temperature fluctuations in GEHRs}

Ionic abundances from CELs often differ from those derived from RLs in ionized nebulae. 
\citet{est98, est99a, est99b} have found that the difference is between 
0.1 and 0.3 dex in three of the brightest Galactic \ion{H}{2} regions. Moreover, 
\citet{tsa03} obtain similar differences for the Orion nebula and M17 and  between 0.30 and 0.69  dex for three Magellanic 
Clouds \ion{H}{2} regions.
In the case of planetary nebulae (PNe) the differences vary from one 
object to another and can be as large as a factor of ten for some PNe 
\citep{rol94, pei95, liu02}. This discrepancy between the abundances obtained 
from CELs and RLs is a crucial problem because most of our knowledge of 
chemical abundances in distant objects -specially in extragalactic ones- 
comes from determinations based on the intensity of CELs.

\citet{tor80} proposed that the abundance discrepancy between calculations 
based on CELs and RLs may be produced by the presence of spatial 
fluctuations of the electron temperature in the nebulae, parametrized by $t^2$ 
\citep{pei67}. This is due to the different dependence of the emissivity of 
CELs and RLs on electron temperature, $T_e$. While the ratio of two RLs is 
almost 
independent of $T_e$, the ratio of a CEL with respect to a RL is strongly 
dependent 
on it.

Another argument in favor of the presence of temperature variations in
gaseous nebulae is the determination of $t^2$ and $T_0$ based on
$T$([\ion{O}{3}]) and
$T$(Bac). Some of the best determinations of $T$(Bac) in planetary nebulae 
are those by  
Peimbert (1971), Liu \& Danziger (1993), Liu et al. (1995, 2000,
2001). Liu et al.(2001), presented a good correlation
between $Log({\rm O}^{++}/{\rm H}^+)_{RL}/({\rm O}^{++}/{\rm H}^+)_{CEL}$ and
$T$([\ion{O}{3}])--$T$(Bac) and mention that this correlation strongly supports
the idea that the temperature variations are real, moreover
Torres-Peimbert \& Peimbert (2003) find that there is good agreemnet between 
the  $t^2$ values determined from the CELs/Rls values and those determined
from $T$([\ion{O}{3}]) and $T$(Bac) for three of the planetary nebulae with the 
largest $t^2$ values in the literature.

Torres-Peimbert and Peimbert(2003) present a review on tempertaure
variations in planetary nebulae and discuss seven possible 
causes that could explain temperature variations in a given object. One of the
possible explanations could be due to the 
presence of a small mass of extremely cold ($T_e$ $\sim$ 10$^3$ K), metal 
and helium-rich inclusions embedded in diffuse material of ``normal'' 
temperature and abundances (see Liu 2002, 2003; Pequignot et al. 2002, 2003)
This mechanism is not important in \ion{H}{2} regions because they are 
not directly produced by stellar ejecta. Esteban(2002) reviews indirect
evidence that indicates the presence of temperature variations in \ion{H}{2} 
regions.

Assuming the validity of the temperature fluctuations paradigm, the comparison of the abundances determined from both kinds 
of lines for a given ion should provide an estimation of $t^2$. In Table 6 we compare 
the O$^{++}$/H$^+$ and C$^{++}$/H$^+$ ratios obtained from  
CELs and RLs and the $t^2$ values that produce the agreement between 
both. 
For the O$^{++}$/H$^+$ ratios, the values obtained from RLs are always 
about 0.2 to 0.35 dex larger than those derived from CELs measured in the same 
spectra, the associated $t^2$ parameter is between 0.027 and 0.128. For
the  C$^{++}$/H$^+$ ratios the comparison between values obtained 
from CELs and RLs is not so direct. The brightest CELs of C$^{++}$ are 
[\ion{C}{3}] 
1907 $+$ \ion{C}{3}]  1909\mbox{\AA}\ , which are in the UV and have to be measured from space 
observations. \citet{gar99} have observed 
NGC~5461 and NGC~5471 with the $FOS$ spectrograph at the $HST$ deriving 
their C$^{++}$ abundance. The C$^{++}$/H$^+$ values included in Table 6 
are the higher and lower limits computed by \citet{gar99} 
because of the uncertainty in the choice of UV reddening function from their 
data.  \citet{pei86} obtained 
$IUE$ measurements of the [\ion{C}{3}] 1907 $+$ \ion{C}{3}]  1909\mbox{\AA}\ 
lines for NGC~2363 and derived its C$^{++}$/H$^+$ ratio, which  is 
also included in Table 6. For NGC~5471 and NGC~2363 the 
values of 
$t^2$ obtained from the comparison of the C$^{++}$ abundances are consistent 
with those obtained from O$^{++}$/H$^+$. For NGC~5461 only the  
C$^{++}$/H$^+$ ratio derived assuming $R_V$ = 5 give a $t^2$ consistent with that
estimated from the comparison of O$^{++}$ abundances. 

The values of $t^2$ obtained for bright Galactic \ion{H}{2} 
regions by \citet{est98, est99a, est99b} are also included in Table 6 for 
comparison. It is important to remark that the C$^{++}$ abundance of the 
two slit positions observed for M17 by \citet{est99b} are now corrected 
due to an error present in that paper [12+log(C$^{++}$/H$^+$) is actually 
8.57 and 8.69 for slit positions 3 and 14 and not 8.76 and 
8.88 as it was indicated in that paper] . In the cases of NGC~604 and NGC~5461, 
the magnitude 
of the fluctuations is modest and of the same order than those determined
for the Galactic objects. Only the GEHRs of lower metallicity --but also 
higher observational uncertainties: NGC~5471 
and NGC~2363-- show larger values of $t^2$. Observations of higher accuracy
are needed to test this result.

The presence of $t^2$ in GEHRs has been a secular matter of discussion 
without a well established observational basis. The first indirect 
evidence reported was the significant differences between the O/H ratios  
obtained using the $R_{23}$ empirical calibration based on models and on 
observations. The differences are typically in the 0.2$-$0.4 dex range. 
Several authors suggested that such differences could be due to the 
presence of temperature inhomogeneities over the observed volume \citep{cam88,tor89,mcg91}. 
It is worth noting also that \citet{pei91} indicate 
that, in the presence of processes increasing the intensity of [O~III] 
4363 \mbox{\AA}\ line (i. e. shocks) the [O~III] 5007 \mbox{\AA}\ and
[O~II] 3727 \mbox{\AA}\ lines, and consequently $R_{23}$, are almost unaffected. 

There are few estimations of $t^2$ available for extragalactic objects. 
\citet{gon94} find large values of $t^2$=0.064-0.098 for NGC~2363 from the 
comparison of $T$([O~III]) and $T$(Pac), values not too different from
our results. On the other hand, \citet{ter96} 
obtain $t^2$$\sim$0.00 for NGC~604 making use of the same method. 
However, the derivation of $t^2$ from the comparison of $T$([O~III]) 
and $T$(Pac) --or $T$(Bac)-- has 
a large intrinsic uncertainty. 
\citet{lur99} have computed photoionization models for NGC~2363 
and compare them with optical observational data. They are able to 
reproduce the observed emission-line spectrum and other properties of 
NGC~2363 only if they assume a metallicity 2.5 times higher than usually 
adopted and derived from the intensity of CELs. \citet{lur99} 
propose that the presence of temperature fluctuations can explain 
this discrepancy. It is remarkable that our results are in complete agreement 
with this suggestion. In fact, if we take the O/H abundance as reference 
of the metallicity and that  O/H is largely O$^{++}$/H$^+$ in NGC~2363, 
our O$^{++}$ abundance obtained from RLs is about 2.2 larger than the 
value we obtain from CELs. Following the conclusions by \citet{lur99}, with 
this corrected metallicity for NGC~2363 it is possible: a) to reconcile the 
presence of WR stars with the current models of stellar evolution; b) 
reproduce most of the characteristics of the observed spectrum of the 
nebula; c) obtain a better agreement with the metallicity estimated making 
use of the $R_{23}$ parameter. 

Another estimation of $t^2$ in an extragalactic \ion{H}{2} region has been  
obtained by \citet{pei00} for  NGC~346 who find $t^2$=0.022 from the 
comparison of $T$([O~III]) and $T$(Bac) and a self-consistent determination 
of $T$(He~II). Additional indications of the possible presence of $t^2$ in 
GEHR and starburst 
galaxies have been given by \citet{est95} and \citet{per97}. \citet{est95} 
studied the chemical enrichment produced by massive stars in 
Wolf-Rayet galaxies, finding that the apparent abnormal position of some of 
these objects in O vs. $Y$ and N vs. $Y$ diagrams can be explained by the effect 
of large $t^2$ in the ionized gas associated with the intense star-forming 
bursts. On the other hand, \citet{per97} has investigated the change of the 
ionization structure during the first Myrs of the evolution of a starburst. 
He has shown that large 
temperature fluctuations arise naturally in homogeneous gaseous spheres 
when the spectral energy distribution of the ionizing cluster hardens at 
about 3 Myr, coinciding with the onset of the WR phase. 

\citet{sta00} has discussed the problems that 
detailed photoionization models face to reproduce the observed [O~III] 
$\lambda$4363/5007 ratio of several GEHRs classified as Wolf-Rayet galaxies. 
The models always predict lower ratios than observed. In the 
case of I Zw 18, \citet{sta99} obtain a temperature discrepancy of the 
order of 30\%, which is a rather large value. \citet{sta00} considers that 
the classical photoionization models fail and that temperature fluctuations 
produced by an unknown process could be present in these objects.

\section{Helium abundances}
The 
He/H value in an  \ion{H}{2}  region is derived from recombination lines that
are proportional to $N_eN(X^{+i})$, therefore  the total
He/H value can be obtained from:

\begin{equation}
\frac{N ({\rm He})}{N ({\rm H})}=
\frac {\int{N_e N({\rm He}^0) dV} + \int{N_e N({\rm He}^+) dV} + 
\int{N_e N({\rm He}^{++})dV}}
{\int{N_e N({\rm H}^0) dV} + \int{N_e N({\rm H}^+) dV}}
\label{eHeH}
,\end{equation}

\noindent in particular it is important to include the $N_e$ factors when there are neutral
atoms in regions of appreciable electron density. Equation (1) can be approximated by: 

\begin{equation}
\frac{N ({\rm He})}{N ({\rm H})} 					
= ICF({\rm He})
\frac {N({\rm He}^+) + N({\rm He}^{++})}
{N({\rm H}^+)}
\label{eICF}
.\end{equation}

To obtain He$^+$/H$^+$ values we need a set of effective recombination
coefficients for the He and H lines, the contribution due to collisional
excitation to the helium line intensities, and an estimate of the optical depth
effects for the helium lines. The recombination coefficients that we used were
those by \citet{hum87} and \citet{sto95} for H, and \citet{smi96} for He. The collisional
contribution to the helium lines was estimated from \citet{kin95} and \citet*{ben99}. 
The collisional contribution to the hydrogen lines was not taken into
account. The optical
depth effects in the triplet lines were estimated from the computations by
\citet{rob68}.

In Table 7 we present the He$^+$/H$^+$ values derived from six helium lines 
assuming no temperature
variations; consequently  $t^2 = 0.00$ and $T_e$([\ion{He}{2}]) is given by the observed 
$T_e$([\ion{O}{3}]) value. To estimate the optical depth of
the triplet lines -represented by $\tau(3889)$- and $N_e$(He~{\sc{ii}}) we made use of Table 3 
and the intensity of \ion{He}{1} 7065 \mbox{\AA}, which is very sensitive to both quantities;
fortunately, the other helium line intensities are considerably less sensitive
to $N_e$(He~{\sc{ii}}) and $\tau(3889)$. The $<$He$^+$/H$^+>$ value is based on the other
helium lines excluding \ion{He}{1} 7065 \mbox{\AA}, with the exception of NGC~2363 where  
\ion{He}{1} 5876 \mbox{\AA}\ was also excluded due to an uncertain reddening correction.
The He$^{++}$/H$^+$ values presented in Table 7 were obtained from the 
\ion{He}{2} 4686 \mbox{\AA}\ to H($\beta$) ratio using line emissivities calculated by \citet{hum87} and \citet{sto95}.
The He$^{0}$/H$^+$ and $ICF$(He) values presented in Table 7 were estimated from CLOUDY
\citep{pei02,lur02}.

Similarly in Table 8 we present the He$^+$/H$^+$ values derived from six helium lines 
assuming  temperature variations; consequently $t^2 > 0.00$ and  $T_e$([\ion{He}{2}]) is smaller 
than the observed $T_e$([\ion{O}{3}]) value \citep{pea02}.

\section{Heavy element abundances}

To derive the total abundances we need to adopt a $t^2$ value and to 
correct for unseen ionization stages by using ionization correction factors. 
The values of $t^2$ adopted for each object are those obtained from the 
comparison of the O$^{++}$/H$^+$ ratios obtained from CELs and RLs, 
which are given in Table 6. 
To correct the ionic abundances derived for the effect of 
temperature fluctuations we have made use of the expressions given by 
\citet{pei69}, some  reference photoionization models by \citet{sta90} (model C2C1 for NGC~604; C2B1 for NGC~5461; the average of C2D1 and D2D1 for NGC~5471; D4E1 for NGC~2363), and the most recent atomic parameters and line emissivities 
available for the different ions used. Since we do not observe [\ion{O}{2}] 
lines in our spectra,  we have derived the O$^+$/H$^+$ ratio from the 
intensity of [\ion{O}{2}] 3727 \mbox{\AA}\ doublet reported by other authors 
(references given in Table 9) 
for the brightest zones of each nebulae and assuming 
our values for the physical conditions. The uncorrected and corrected 
ionic and total abundances are presented in Table 9. 

The absence or faintness of the \ion{He}{2} 4686 \mbox{\AA}\ line (and the low 
He$^{++}$/H$^+$ ratio associated) and the similarity between the ionization 
potentials of He$^+$ and O$^{++}$ implies the absence of significant 
O$^{3+}$ in the objects. Therefore, to obtain the total oxygen abundance 
we have simply added the O$^+$/H$^+$ and O$^{++}$/H$^+$ ratios. 

To derive the N abundance we have used the usual $ICF$ based on 
the similarity between the ionization potential of N$^+$ and O$^+$ 
\citep{pei69} for all the objects. For these objects the models of 
\citet{sta90}
give values of the $ICFs$ similar to the $ICFs$ by Peimbert
and Costero. Alternatively the models based on CLOUDY give $ICFs$ somewhat
larger \citep{rel02,lur02}.

To derive the C abundance we only have direct determinations of the 
C$^{++}$/H$^+$ ratios therefore we have adopted the following equation

\begin{equation}
 \frac{N ({\rm C})}{N ({\rm H})} 					
=\frac{N ({\rm O})}{N ({\rm H})}
 \frac{N ({\rm C})}{N ({\rm O})} 
=\frac{N ({\rm O})}{N ({\rm H})}
ICF({\rm C^{++}}/{\rm O^{++}})
\frac{N ({\rm C^{++}})}{N ({\rm O^{++}})}
\label{CICF}
,\end{equation}

where the C$^{++}$/O$^{++}$ ratios are determined from RLs.
The $ICF$(C$^{++}$/O$^{++}$) values have been estimated from
the models by \citet{sta90} and \citet{gar99} and fortunately
are close to one.

We have measured CELs from two ionization stages of S: S$^+$ and S$^{++}$. 
Taking into account the ionization degree of the objects, an $ICF$ for the 
presence of S$^{3+}$ has to be considered in all cases. We have adopted the 
$ICF$ given by the reference models, and the expected amount of  
S$^{3+}$ results to be important in all 
cases, ranging from 1.45 for NGC~5461 to 5.15 for NGC~2363. 
Taking into account these high $ICF$s we consider that the total S 
abundance we have derived is uncertain, specially for the objects of 
higher ionization degree: NGC~5471 and NGC~2363. 

For Ar we have determinations of Ar$^{++}$/H$^+$ and Ar$^{3+}$/H$^+$. 
We do not expect significant contributions of Ar$^{4+}$ but some Ar$^+$ 
may be present. We have adopted the $ICF$(Ar$^{++}$+Ar$^{3+}$) given by 
the reference photoionization models of Stasi\'nska for all the objects. 
The $ICFs$ are always very small being the largest that of NGC 5461, 
which amounts to 1.06. 

We have measured lines of only one stage of ionization (Fe$^{++}$)  
for NGC~604 and NGC~5461. In the case of NGC~5471 and NGC~2363 
we have also detected lines of Fe$^{3+}$. The contribution of Fe$^{3+}$ 
is expected to be very important. On the other hand, some small 
contribution of  Fe$^+$ should be present, mainly in the nebulae of lower ionization degree. For all the objects, we have estimated the Fe/H ratio 
from the derived Fe$^{++}$ abundance and adopting the $ICF$(Fe$^{++}$) given by the reference photoionization 
models. These models give very large values of that factor ranging from 
6.2 to 41 (for NGC~604 and 
NGC~2363 respectively). We have compared with the 
$ICF$ given by the relation between the Fe$^{++}$ abundance 
and O$^+$/O$^{++}$ obtained by \citet{rod96} from the models of 
\citet{gru92}, finding that the differences with our adopted values of 
the $ICF$(Fe$^{++}$) are only about 0.1 dex.  In the cases of NGC~5471 
and NGC~2363 we have also derived the gas-phase Fe/H ratio simply adding 
the Fe$^{++}$/H$^+$ and Fe$^{3+}$/H$^+$ ratios, which are expected to contain about the 97\% of the total Fe abundance in the gas-phase 
according to the reference models. The Fe/H ratio obtained assuming $ICF$s 
are in both cases much larger than the direct determination (0.35 dex for 
NGC~5471 and 0.65 for NGC~2363). This result was also obtained by 
\citet{rub97} in the case of the Orion nebula, where their determination 
of the Fe/H ratio from the intensity of [\ion{Fe}{4}] 2837 \mbox{\AA}\ line are much 
lower than the values obtained from the Fe$^{++}$/H$^+$ ratios and assuming 
an $ICF$ from detailed photoionization models. We agree with the 
suggestion of \citet{rub97} that a reexamination of Fe$^{3+}$ atomic 
data and perhaps improving \ion{H}{2} region modeling would help 
to solve this problem. 

\section{Discussion}

\subsection{The O/H and C/H gradients in M101}

The determination of the O/H and C/H ratios in NGC~5461 and NGC~5471 
allows to derive the radial abundance gradients of these two elements 
in M101. The galactocentric distance of both \ion{H}{2} regions is very 
different. NGC~5461 is located at 11.1 kpc and NGC~5471 at 26.2 kpc assuming a 
distance of 7.5 Mpc for M101 \citep{kel96}. 

The oxygen abundance gradient has been derived for M101 
by many authors \citep{eva86,vil92,hen95,ken96,pil01}. Our value of 
$-$0.038 dex kpc$^{-1}$ has been computed from the O/H ratios obtained from 
RLs (using CELs and $t^2=0.00$ the slope becomes $-$0.035 dex kpc$^{-1}$) and 
it is consistent with 
previous determinations ($-$0.044: Vila-Costas and Edmunds 1992; 
$-$0.040 to $-$0.050: Kennicutt and Garnett 1992; $-$0.028: 
Pilyugin 2001) although it is rather uncertain due to the large observational 
error of the O/H ratio of NGC~5471.

\citet{gar99} based on $HST$ UV observations determined the presence of 
C/H and C/O gradients in external galaxies for the first time. For M101, 
those authors estimate two values of the C/H and two values for
the C/O gradient depending on the $R_V$ value adopted (see Table 10).
Our values of the C/H and C/O gradients derived from RLs
are independent of the $R_V$ value adopted and are also presented in
Table 10.

The values of C/O obtained for the objects in this paper based on RLs
are similar or larger than previous 
determinations based on the intensity of CELs. For NGC~5461, 
our value is similar to that obtained by \citet{gar99} assuming 
$R_V$ = 5 ($-$0.37 dex) but smaller than their C/O ratio for $R_V$ = 3.1 
($-$0.03 dex). For NGC~5471 and NGC~2363 our C/O ratios 
are 0.22 and 0.05 dex higher that those obtained by other 
authors \citep{gar99,pei86}.

The value of the C/H gradient and the C/O ratio derived by us
for NGC 5461 favor the results for $R_V$ = 5.0 of \citet{gar99}.
Our C/H value for NGC 5471 is very uncertain, therefore for
the following discussion we will consider the C/O gradient of
\citet{gar99} for $R_V$ = 5.0.

\subsection{The C/O ratio and chemical evolution}

The exact value of the C/O gradient in a spiral galaxy is 
a strong constraint for models of chemical evolution and 
the star formation history. The evolution of C/O with O/H is very sensitive 
to the star formation/enrichment timescale because the different initial 
mass range of the main producers of both elements. 
 
The O/H gradient for the Milky Way 
obtained by \citet{est99b} based on RLs is rather similar to the value found 
for M101 and it has been also included in Table 10 for comparison. 
The slope of the Galactic 
C/H gradient included in Table 10 is a revision of the 
value given by \citet{est99b} in their Table 13. This new slope has 
been re-calculated considering the recombination coefficients by
\citet{dav00} and an error in the C$^{++}$/H$^+$ ratio of M~17 present
in that paper (see Sect. 6).  

A general correlation between C/O and O/H was obtained by \citet{gar95} 
from UV observations of \ion{H}{2} regions of dwarf irregular galaxies. 
This trend was later confirmed by \citet{gar99} from similar 
observations of six \ion{H}{2} regions in the spiral galaxies M101 and 
NGC~2403. We have found a similar behavior in our data, which are 
shown in Figure 5 along with the values for the Orion nebula, M8 and M17 
\citep{est98, est99a, est99b}, and the Sun 
\citep{all01,hol01}. This graph is relevant because all the nebular 
data included are abundances derived from recombination lines, 
almost insensitive to potential electron temperature variations 
inside the nebulae.

The presence of C/O gradients in the Milky Way and M101 is a 
reflection of this correlation between C/O and metallicity. The behavior 
of C/O vs. O/H can be approximated by a power law
\begin{equation}
{\rm log(C/O)} = a + b~{\rm log(O/H)}
\end{equation}
We have performed a regression analysis of the nebular data included in 
Figure 5 finding $b$ = 0.56 $\pm$ 0.10 (by considering only the four
GEHRS we find $b$ = 0.30 $\pm$ 0.15) . This value of the slope is 
consistent with those obtained by \citet{gar97} from a compilation of data 
for dwarf irregular galaxies (0.41$\pm$0.07) and the larger compilation 
by \citet{hen99} that includes irregular and spiral galaxies (slope between 
0.48 and 0.59). This behaviour gives important insights into the origin and 
evolution of both elements. Simple chemical evolution models with 
instantaneous recycling predict a C/O = constant if both C and O are pure 
primary elements, and C/O $\propto$ O/H if O is primary and C 
secondary. As \citet{gar99} have pointed out, since only primary sources 
of C are known to exist, the observed behavior of C/O vs. O/H indicates 
that the instantaneous recycling approximation does not hold for both 
elements and/or the yield of C varies with respect to O. 
In chemical evolution models that assume infall of primordial gas without 
galactic winds or radial outflows and a star formation rate 
proportional to $\sigma_{gas}^x \sigma_{tot}^{x-1}$ (where $ \sigma_{gas}$
is the surface gas density, $ \sigma_{tot}$ is the total surface mass 
density, and $x$ is constant in time and space), the predicted C/O ratio 
depends mainly on the stellar yields and the initial mass function. 

Based on chemical 
evolution models for the Galactic disk \citet{car00} has studied the C and O 
gradients assuming different sets of stellar yields dependent on 
metallicity.  This author has shown that the observed C/O vs. O/H 
relation --as 
well as the presence of negative C/O gradients in spiral galaxies-- can be 
explained with the metallicity-dependent yields for C and O for massive 
stars with radiatively driven mass loss derived by \citet{mae92}. In Figure 5 
we have also plotted the prediction of selected models (those that reproduce 
the observed relation) obtained by \citet{car00}. It can be seen that the 
position of the Galactic objects is well reproduced by the models. 
However, as \citet{gar99} have pointed out changes in the assumed 
star formation law through its dependence on the power of the 
gas surface density can change the curves, mainly shifting the curves 
in the C/O axis. Therefore, models for the solar vicinity do not necessarily  
reproduce correctly the chemical evolution of galaxies with different 
gas consumption timescales.
 
\subsection{The $\Delta Y$/$\Delta O$ and $\Delta Y$/$\Delta Z$ ratios 
and chemical evolution}

Additional parameters that can be used to constrain models of galactic
chemical evolution are $\Delta Y/\Delta O$ and $\Delta Y$/$\Delta Z$; these 
values are obtained by
assuming that the galaxies form with the primordial helium abundance
and no heavy elements. The best objects to obtain
this parameter are NGC 604 and NGC 5461 due to their high $O$ and
$Y$ abundances. 

In Table 11 we present the helium, oxygen and heavy elements by mass
($Y$, $O$, and $Z$) for NGC 604 and NGC 5461. To derive the $O$ values 
we have added 0.08 dex to the gaseous values to take into account the 
fraction of heavy elements tied up in dust grains.
To derive the heavy element content by mass of NGC 604 and NGC 5461 we
made the following considerations: a) we added 0.08 dex and 0.10 dex to
the $O$ and $C$ gaseous abundances respectively to take into account the fraction 
of these atoms
tied up in dust grains \citep{est98}; b) we assumed that $C+N+O$ were
representative of the total $Z$ value and based on the solar abundances by \citet{gre98}
and by \citet{hol01} we obtained that $C+N+O$ made up 69\% of the total $Z$ value,

Also in Table 11 we include the $Y$, $O$ and $Z$
values obtained by \citet{pei02} based on the line intensities by
\citet{est99b} for M17, the galactic  \ion{H}{2}  region with the best
helium abundance determination. To derive the total $Z$ values these
authors assumed that  $C$, $N$, $O$ and $Ne$ were representative of the $Z$ value,
and based on the solar abundances by \citet{gre98} and \citet{hol01} they
found that these four elements amount to 79\% of the total $Z$ value.

We have assumed the primordial helium abundances 
by mass recently obtained by \citet{pea02} of $Y_p$(+Hc) = 0.2475 $\pm$0.0025
for $t^2$ = 0 and $Y_p$(+Hc) = 0.2384$\pm$0.0025 for $t^2$ $>$ 0,
that were obtained considering the collisional excitation of the Balmer lines
(+Hc). In Table 12 
we have included the values of $\Delta Y$/$\Delta O$ that we obtain assuming 
$t^2$ = 0 and $t^2$ $>$ 0.

{From} the 
data shown in Table 12 it is remarkable that we obtain fairly similar values 
of $\Delta Y/\Delta O$ for $t^2$ = 0, which range  from 4.6 to 6.6. 
These ratios for $t^2$ = 0 are larger than those obtained by chemical 
evolution models. For the galactic disk at the solar 
vicinity \citet{chi97} find $\Delta Y/\Delta O$ = 3.15, \citet{cop97} 
derived values in the 2.4 to 3.4 range, and \citet{car00} obtains 
$\Delta Y/\Delta O$ ratios between 2.9 and 4.6, while closed box models for 
irregular galaxies by \citet{car95} and \citet {car99} give values in the 2.95
to 4.2 range. 
In Table 11 we can see that the $\Delta Y/\Delta O$ ratios for $t^2$ $>$ 0
are in the 2.55 to 4 range in good
agreement with the galactic chemical evolution models.

Similarly the models of chemical evolution of the Galaxy mentioned in the
previous paragraph predict $\Delta Y/\Delta Z$ values in the 1.2--2 range.  Again
the values derived from observations under the assumption that  $t^2$ $>$ 0
are in better agreement with the models than those derived under the assumption
that $t^2 = 0$ (see Table 12).

\section{Conclusions}

Recombination lines of C II and O II are detected by the first time in 
some of the most interesting GEHRs of the Northern hemisphere. The comparison of ionic abundances obtained from these lines with those obtained from CELs suggests the presence of temperature variations
inside these nebulae.

[\ion{Fe}{4}] optical  lines are detected in NGC 2363 and NGC 5471, this is 
the clearest identification of these lines in \ion{H}{2} regions.

The presence of values of $t^2$ $>$ 0 in GEHRs has important 
consequences in various  fields of astrophysics, mainly because most of 
our knowledge about the chemical content of extragalactic objects comes 
from the spectra of GEHRs. The usual assumption of $t^2 = 0$ leads
to underestimate the heavy element abundances derived from
CELs. Moreover the usual assumption of $t^2 = 0$ leads
to underestimate $Y_p$ the primordial helium abundance \citep{pei00,pei02}.
 
{From} the RLs of C and O we confirm the existence of
C/H and O/H gradients in M101 derived from CELs.

We find a poor agreement between the predicted
$\Delta Y$/$\Delta O$ and $\Delta Y$/$\Delta Z$ values by galactic chemical 
evolution models
and those derived from observations under the assumption that $t^2 = 0$. 
Alternatively we find a good agreement between the predicted values and
those derived from the RLs of C and O and the CELs 
of N, Ne, S, Ar, and Fe under the assumption that $t^2$ $>$ 0.

\acknowledgments

We would like to acknowledge Leticia Carigi for her help with aspects 
concerning chemical evolution models and Antonio Peimbert for his help on the
determination of the helium abundances. We are grateful to an anonymous referee  
for his/her very useful comments. C. E. would like to thank all the members 
of the Instituto de Astronom\'\i a, 
UNAM, for their warm hospitality during his stays in Mexico. 
Financial support has been provided through grant PB97-1435-C02-01 
from DGES, Spain.

\clearpage


\begin{deluxetable}{lccccccccc}
\tabletypesize{\footnotesize}
\tablenum{1}
\tablecaption{Summary of observations \label{tbl-1}}
\tablewidth{0pt}
\tablehead{
\colhead{} & \colhead{} & \colhead{} & \colhead{} & \colhead{} & \colhead{} & 
\colhead{Spectral} & \colhead{Spatial} & \colhead{Exposure} \\
\colhead{} & \colhead{R.A.} & \colhead{Dec.} & P.A. & \colhead{Size} & \colhead{$\Delta\lambda$} & \colhead{resol.} & \colhead{scale} & 
\colhead{time} \\
\colhead{Object} & \colhead{(hh mm ss)} & 
\colhead{($^\circ$ ${\tt '}$ ${\tt ''}$)} & \colhead{($^\circ$)} & 
\colhead{(arcsec$^2$)} &\colhead{($\rm\mbox{\AA}$)} & 
\colhead{($\rm\mbox{\AA}$ pix$^{-1}$)} &  
\colhead{(arcsec pix$^{-1}$)} & \colhead{(s)} & 
}
\startdata
NGC 604 & 01 34 32.7 & +30 47 02 & 305 & 12.7 & 4230$-$5060 & 0.225 & 0.19 & 9000   \\
& & & & & 5720$-$7200 & 1.47 & 0.36 & 9000  \\ 
& & & & & & & & 300 \\
NGC 5461 & 14 03 39.8 & +54 18 51 & 240 & 4.6 & 
4230$-$5060 & 0.225 & 0.19 & 9600 \\
& & & & & 5720$-$7200 & 1.47 & 0.36 & 9600  \\
& & & & & & & & 1200 \\
NGC 5471 & 14 04 27.7 & +54 23 43 & 
310 & 4.4 & 4230$-$5060 & 0.225 & 0.19 & 12600 \\
& & & & & 5720$-$7200 & 1.47 & 0.36 & 12600 \\
NGC 2363 & 07 28 43.8 & +69 11 15 & 80 & 4.9 & 4230$-$5060 & 0.225 & 
0.19 &25200 \\
& & & & & 5720$-$7200 & 1.47 & 0.36 & 25200  \\
& & & & & & & & 300 \\
\enddata
\end{deluxetable}

\clearpage

\begin{deluxetable}{lccccccccccc}
\tabletypesize{\tiny}
\setlength{\tabcolsep}{0.07in}
\tablecaption{line intensities \label{tbl-2}}
\tablehead{
\colhead{} & \colhead{} & \colhead{} & \colhead{} & \multicolumn{2}{c}{NGC 604} & \multicolumn{2}{c}{NGC 5461} & 
\multicolumn{2}{c}{NGC 5471} & \multicolumn{2}{c}{NGC 2363} \\
\colhead{$\lambda$$_0$} & \colhead{} & \colhead{} & \colhead{} & \colhead{$\lambda$} & \colhead{}& \colhead{$\lambda$} & \colhead{} & \colhead{$\lambda$} &  \colhead{} & \colhead{$\lambda$} & \colhead{} \\
\colhead{($\rm\mbox{\AA}$)} & \colhead{Ion} & \colhead{Mult.} & \colhead{f($\lambda$)} & \colhead{($\rm\mbox{\AA}$)} & \colhead{I($\lambda$)} &  \colhead{($\rm\mbox{\AA}$)} & \colhead{I($\lambda$)} & \colhead{($\rm\mbox{\AA}$)} & \colhead{I($\lambda$)} & \colhead{($\rm\mbox{\AA}$)} & \colhead{I($\lambda$)}
}
\startdata
4267.16 &  \ion{C}{2} & (6) & 0.141 & 4267.17 & 0.20$\pm$0.03 & 4267.11 & 0.18$\pm$0.04 & 4267.24 & 0.05: & 4267.07 & 0.04: \\
4275.55 &  \ion{O}{2} & (68) & 0.140 & 4275.96 & 0.03: & \nodata & \nodata & \nodata & \nodata &  \nodata & \nodata \\
4276.75 & \ion{O}{2} & (67) & 0.139 & 4276.64 & 0.02: & \nodata & \nodata & \nodata & \nodata &  \nodata & \nodata \\
4276.83 &  [\ion{Fe}{2}] & (21F)& 0.139 & &  & & & & & &  \\
4287.40 & [\ion{Fe}{2}] & (7F) & 0.138 & 4287.51 &  0.04$\pm$0.01 & 4287.02 &  0.11$\pm$0.03 & 4287.25 & 0.10$\pm$0.03 & 4287.61 &  0.03: \\
4303.61 & \ion{O}{2} & (66) & 0.135 & 4303.93 & 0.09$\pm$0.02 & 4303.93 & 0.07: & \nodata & \nodata & 4303.40 & 0.01: \\
4303.82 & \ion{O}{2} & (53) & 0.135 & & & & & & & & \\
4317.14 & \ion{O}{2} & (2) & 0.132 & 4317.30 & 0.04: &  \nodata & \nodata & \nodata & \nodata & 4317.25 & 0.01: \\
4340.47 & \ion{H}{1} & H$\gamma$ & 0.129 & 4340.49 & 46.6$\pm$0.9  & 4340.45 & 46.5$\pm$0.09 &4340.46 &  47.6$\pm$0.09 & 4340.47 & 47.3$\pm$0.9 \\
4345.56 &  \ion{O}{2} & (2) & 0.128 & 4345.63 & 0.05: & 4345.71 & 0.03: &  \nodata & \nodata &  \nodata & \nodata \\
4349.43 &  \ion{O}{2} & (2) & 0.127 & 4349.34 & 0.05: &  \nodata & \nodata &  \nodata & \nodata &  \nodata & \nodata \\
4359.34 &  [\ion{Fe}{2}] & (7F) & 0.125 &  \nodata & \nodata & 4359.14 & 0.05: &  \nodata & \nodata &  \nodata & \nodata \\
4363.21 &  [\ion{O}{3}] & (2F) & 0.124 & 4363.22 & 0.71$\pm$0.05 & 4363.21 &  1.27$\pm$0.13 & 4363.23 & 10.3$\pm$0.5 &  4363.20 & 15.9$\pm$0.5 \\
4366.89 &  \ion{O}{2} & (2) & 0.123 & 4366.89 & 0.03: &  \nodata & \nodata &  \nodata & \nodata &  \nodata & \nodata \\
4368.25 & \ion{O}{1} & (5) & 0.123 &  \nodata & \nodata &  \nodata & \nodata &  \nodata & \nodata &  4368.04 & 0.05: \\
4387.93 &  \ion{He}{1} & (51) & 0.118 & 4387.94 & 0.60$\pm$0.04 & 4387.98 & 0.56$\pm$0.05 & 4387.91 &  0.52$\pm$0.06 & 4387.92 &  0.52$\pm$0.04 \\
4413.78 & [\ion{Fe}{2}] & (7F) & 0.111 &  \nodata & \nodata & 4413.72 &  0.06: &  \nodata & \nodata & 4413.85 & 0.02: \\ 
4416.27 &  [\ion{Fe}{3}] & (6F) & 0.113 & 4417.16 & 0.02: & \nodata & \nodata &  \nodata & \nodata &   \nodata & \nodata \\
4416.97 &  \ion{O}{2} & (5) & 0.113 & & & & & & & & \\
4437.55 &  \ion{He}{1} & (50) & 0.104 & 4437.40 & 0.07$\pm$0.02 & 4437.32 &  0.06: &  \nodata & \nodata & 4437.52 & 0.08$\pm$0.03 \\ 
4471.48 & \ion{He}{1} & (14) & 0.095 & 4471.52 & 4.38$\pm$0.16 & 4471.51 & 4.44$\pm$0.16 & 4471.51 &  4.15$\pm$0.23 & 4471.50 & 4.07$\pm$0.16 \\
4491.22 & \ion{O}{2} & (86) & 0.092 & 4491.21 & 0.03: &  \nodata & \nodata &  \nodata & \nodata &   \nodata & \nodata \\
4562.60 & \ion{Mg}{1}] & (1) & 0.074 & 4562.62 & 0.06$\pm$0.02 & 4562.40 & 0.07$\pm$0.03 & 4562.47 &  0.19$\pm$0.05 & 4562.57 & 0.09$\pm$0.02 \\
4571.10 & \ion{Mg}{1}] & (1) & 0.072 & 4571.07 & 0.07$\pm$0.02 & 4571.18 & 0.05$\pm$0.02& 4571.01 &  0.16$\pm$0.05 & 4571.03 & 0.06$\pm$0.02 \\
4590.97 &  \ion{O}{2}  & (15) & 0.067 & 4591.02 & 0.03: & 4591.02 & 0.03: &  \nodata & 
\nodata & \nodata & \nodata \\
4638.85 &  \ion{O}{2} & (1) & 0.053 & 4638.83 & 0.08$\pm$0.03 & 4638.67 & 0.07$\pm$0.03 &  \nodata & \nodata & \nodata & \nodata \\
4641.81 &  \ion{O}{2} & (1) & 0.053 & 4641.60 & 0.06$\pm$0.02 & 4641.72 & 0.13$\pm$0.04 & 4641.75 &  0.04: & 4641.64 & 0.05$\pm$0.02 \\
4649.14 &  \ion{O}{2} & (1) & 0.051 & 4649.13 &  0.05$\pm$0.01 & 4650.13 & 0.22$\pm$0.06 &  \nodata & \nodata & 4649.88 & 0.08$\pm$0.02 \\
4650.84 &  \ion{O}{2} & (1) & 0.050 & 4650.87 & 0.08$\pm$0.02 & & &  & & & \\
4658.10 & [\ion{Fe}{3}] & (3F) & 0.048 & 4658.14 & 0.30$\pm$0.03 & 4658.06 & 0.63$\pm$0.08 & 4658.17 &  0.69$\pm$0.10 & 4658.13 & 0.15$\pm$0.03 \\
4661.64 &  \ion{O}{2} & (1) & 0.048 & 4661.52 & 0.06$\pm$0.01 & 4661.46 & 0.05: &  \nodata & \nodata & 4661.83 & 0.04$\pm$0.01 \\
4676.23 &  \ion{O}{2} & (1) & \nodata & \nodata & \nodata & \nodata & \nodata &  \nodata & \nodata & 4676.09 & 0.01: \\
4685.71 & \ion{He}{2} & (1) & 0.043 &  \nodata & \nodata & 4684.67 & 0.14$\pm$0.04 & 4685.83 &  0.57$\pm$0.09 & 4685.94 & 0.41$\pm$ 0.04\\
4701.62 &  [\ion{Fe}{3}] & (3F) & 0.037 & 4701.57 & 0.07$\pm$0.02 & 4701.50 & 0.16$\pm$0.04 & 4701.71 &  0.16: & 4701.53 & 0.03: \\ 
4711.34 &  [\ion{Ar}{4}] & (1F) & 0.036 & 4711.14 & 0.05$\pm$0.01 & \nodata & \nodata & 4711.46 &  1.03$\pm$0.08 & 4711.34 & 2.66$\pm$0.08 \\
4713.14 &  \ion{He}{1} & (12) & 0.036 & 4713.13 & 0.37$\pm$0.03 & 4713.13 & 0.37$\pm$0.06 & 4713.24 &  0.37$\pm$0.07 & 4713.12 & 0.62$\pm$0.05 \\
4740.20 & [\ion{Ar}{4}] & (1F) & 0.029 & 4740.72 & 0.02: & 4740.32 & 0.04: & 4740.31 &  0.80$\pm$0.10 & 4740.18 & 2.14$\pm$0.09 \\
4754.69 &   [\ion{Fe}{3}] & (3F) & 0.026 & 4754.80 & 0.03: & 4754.58 & 0.08$\pm$0.03 & 4754.60 &  0.10: & 4755.03 & 0.03: \\ 
4814.55 & [\ion{Fe}{2}] & (20F) & 0.011 & 4815.22 & 0.05$\pm$0.01 &  \nodata & \nodata &  \nodata & \nodata & \nodata & \nodata \\
4815.51 &  \ion{S}{2} & (9) & 0.011 & & & & & & & & \\
4861.33 &  \ion{H}{1} & H$\beta$ & 0.000 & 4861.31 & 100$\pm$2 & 4861.33 & 100$\pm$2 & 4861.33 &  100$\pm$2 & 4861.30 & 100$\pm$2 \\
4867.95+8.16& [\ion{Fe}{4}] & ($^4G-^4F$) & $-$0.002 & \nodata & \nodata & \nodata & \nodata & \nodata &  \nodata &  4867.71 & 0.02: \\
4881.00 &  [\ion{Fe}{3}] & (2F) & $-$0.005 & 4881.07 & 0.09$\pm$0.02 & 4880.95 & 0.21$\pm$0.05 & 4881.39 &  0.27$\pm$0.08 & 4881.07 & 0.04: \\
4903.07 & [\ion{Fe}{4}] & ($^4G-^4F$) & $-$0.009 & \nodata & \nodata & \nodata & \nodata & 4902.97 & 0.04: & 4902.97 & 0.03: \\
4906.56 &  [\ion{Fe}{4}] & ($^4G-^4F$) & $-$0.010 & \nodata& \nodata& \nodata& \nodata& \nodata& \nodata& 4906.59 & 0.04: \\
4921.93 &  \ion{He}{1} & (48) & $-$0.014 & 4921.90 & 1.23$\pm$0.05 & 4921.94 & 1.23$\pm$0.09 & 4921.94 & 1.01$\pm$0.07 & 4921.90 & 1.06$\pm$0.06 \\
4931.23 &  [\ion{O}{3}] & (1F) & $-$0.016 & 4931.17 & 0.02: &  \nodata & \nodata & 4931.34 & 0.09: &  4931.19 & 0.09$\pm$0.03 \\
4958.91 &  [\ion{O}{3}] & (1F) & $-$0.023 & 4958.88 & 78.0$\pm$1.7 & 4958.95 & 112$\pm$2 & 4958.97 & 209$\pm$4 & 4958.87 & 244$\pm$5 \\
4985.90 &   [\ion{Fe}{3}] & (2F) & $-$0.029 & 4985.84 & 0.27$\pm$0.04 & 4985.60 & 0.57$\pm$0.11 & 4985.79  & 0.41$\pm$0.08 & 4985.89 & 0.10$\pm$0.03 \\
5006.84 & [\ion{O}{3}] & (1F) & $-$0.034 & 5006.81 & 250$\pm$4 & 5006.85 & 352$\pm$7 & 5006.88 &  640$\pm$10 & 5006.80 & 729$\pm$14 \\
5015.68 &  \ion{He}{1} & (4) & $-$0.036 & 5015.65 & 2.12$\pm$0.11 & 5015.65 & 1.98$\pm$0.14 & 5015.64 & 1.44$\pm$0.09 &  5015.64 & 1.65$\pm$0.12 \\
5754.64 &   [\ion{N}{2}] & (3F) & $-$0.191 & 5753.86 & 0.28$\pm$0.04 & 5755.59 &  0.35$\pm$0.04 & 5753.29 & 0.14$\pm$0.02 & 5755.59 & 0.07$\pm$0.01 \\
5875.67 &  \ion{He}{1} & (11) & $-$0.216 & 5874.81 & 13.1$\pm$0.4 & 5876.28 & 12.7$\pm$0.04 & 5874.48 & 11.8$\pm$0.5 &  5876.28 & 12.3$\pm$0.2 \\
5957.67 &  \ion{Si}{2} & (4) & $-$0.229 & 5957.27 & 0.06: & 5957.73 & 0.06$\pm$0.02 & 5956.31 & 0.03: & 5957.79 & 0.05: \\
5978.93 &  \ion{Si}{2} & (4) &  $-$0.233 & 5977.69 & 0.06: & 5979.06 & 0.06$\pm$0.02 & 5977.19 &  0.03: & 5979.07 & 0.03: \\
6046.44 & \ion{O}{1} & (22) & $-$0.244 &  \nodata & \nodata &  \nodata & \nodata & 6043.95 &  0.08: & 6046.53 & 0.02: \\
6101.83 & [\ion{K}{4} ]& (1F) & $-$0.253 &  \nodata & \nodata &  \nodata & \nodata & 6099.18 &  0.06: & 6102.31 & 0.14$\pm$0.02 \\
6300.30 &   [\ion{O}{1}] & (1F) & $-$0.285 & 6299.16 & 0.27$\pm$0.02 & 6300.93 & 1.31$\pm$0.07 & 6298.57 &  2.64$\pm$0.18 & 6300.70 & 0.88$\pm$0.06 \\
6312.10 &  [\ion{S}{3}] & (3F) & $-$0.287 & 6310.05 & 1.46$\pm$0.06 & 6312.82 & 1.37$\pm$0.07 & 6310.49 &  1.64$\pm$0.16 & 6312.47 &1.29$\pm$0.06 \\
6347.09 &  \ion{Si}{2} & (2) & $-$0.291 & 6345.56 &  0.06: & 6348.20 & 0.15$\pm$0.04 & 6345.20 &  0.08: & 6348.03 & 0.07$\pm$0.02 \\
6363.78 &  [\ion{O}{1}] & (1F) & $-$0.294 & 6362.90 & 0.33$\pm$0.03 & 6364.43 & 0.39$\pm$0.06 & 6362.02 &   0.92$\pm$0.14 & 6364.17 & 0.31$\pm$0.05 \\
6371.36 &  \ion{Si}{2} & (2) & $-$0.295 & 6370.60 & 0.09$\pm$0.02 & 6372.37 & 0.09$\pm$0.03 &   \nodata & \nodata &  6370.82 & 0.03: \\
6548.03 &  [\ion{N}{2}] & (1F) & $-$0.321 & 6547.34 & 9.84$\pm$0.40 & 6548.83 & 10.8$\pm$0.3 & 6546.32 &  1.89$\pm$0.19 & 6548.28 & 0.45$\pm$0.05 \\
6562.82 &  \ion{H}{1} & H$\alpha$ & $-$0.323 & 6561.80 & 291$\pm$5 & 6563.48 & 291$\pm$6 & 6561.17 &  278$\pm$5 & 6563.03 & 278$\pm$6 \\
6583.41 & [\ion{N}{2}] & (1F)& $-$0.326 & 6582.37 & 26.3$\pm$0.5 & 6584.13 & 31.2$\pm$0.6 & 6581.78 &  6.23$\pm$0.25 & 6583.67 & 1.51$\pm$0.09 \\
6678.15 &  \ion{He}{1} & (46) & $-$0.338 & 6676.84 & 3.69$\pm$0.15 & 6678.76 & 3.60$\pm$0.18 & 6676.49 &  2.95$\pm$0.24 & 6678.27 & 2.88$\pm$0.09 \\
6716.47 & [\ion{S}{2}] & (2F) &  $-$0.343 & 6715.16 & 14.4$\pm$0.4 & 6717.05 & 11.7$\pm$0.4 & 6714.84 & 8.95$\pm$0.27 & 6716.48 & 3.15$\pm$0.09 \\
6730.85 &  [\ion{S}{2}] & (2F) & $-$0.345 & 6729.60 & 10.7$\pm$0.3 & 6731.46 & 10.0$\pm$0.3 & 6729.22 & 7.26$\pm$0.22 & 6730.85 & 2.71$\pm$0.08 \\
6761.3 & [\ion{Fe}{4}] & ($^4G-^2I$)& $-$0.349 & \nodata & \nodata & \nodata & \nodata & \nodata & \nodata & 6760.87 & 0.01: \\
7065.28 &   \ion{He}{1} & (10) & $-$0.383 & 7062.08 & 2.21$\pm$0.09 & 7066.66 & 3.32$\pm$0.17 & 7064.06 & 2.34$\pm$0.23 & 7063.90 & 3.50$\pm$0.10 \\
7135.79 &  [\ion{Ar}{3}] & (1F) & $-$0.391 & 7131.69 & 10.0$\pm$0.3 & 7137.63 & 11.4$\pm$0.4 & 7134.78 & 6.23$\pm$0.19 & 7133.86 & 4.81$\pm$0.14 \\
\\
$C$(H$\beta$) & & & & \multicolumn{2}{c}{0.27$\pm$0.07} & \multicolumn{2}{c}{0.30$\pm$0.07} & 
\multicolumn{2}{c}{0.00$\pm$0.07} & \multicolumn{2}{c}{0.72$\pm$0.08} \\
\multicolumn{4}{l}{$F$(H$\beta$) (10$^{-14}$erg cm$^{-2}$ s$^{-1}$ arcsec$^{-2}$)} & \multicolumn{2}{c}{2.33} & \multicolumn{2}{c}{2.75} & 
\multicolumn{2}{c}{2.16} & \multicolumn{2}{c}{3.28} \\
\multicolumn{4}{l}{$EW$(H$\beta$) ($\rm\mbox{\AA}$)} & \multicolumn{2}{c}{400} & \multicolumn{2}{c}{190} & 
\multicolumn{2}{c}{225} & \multicolumn{2}{c}{325} \\
\enddata
\end{deluxetable}

\clearpage

\begin{deluxetable}{lcccc}
\tablewidth{0pt}
\tablenum{3}
\tablecaption{Physical conditions \label{tbl-3}}
\tablehead{
\colhead{Parameter} & \colhead{NGC 604} & \colhead{NGC 5461} & 
\colhead{NGC 5471} & \colhead{NGC 2363} }
\startdata
$N_e$([\ion{S}{2}]) (cm$^{-3}$) & $\leq$100 & 300$\pm$70 & 
220$\pm$70 & 360$\pm$90 \\
$N_e$([\ion{Ar}{4}]) (cm$^{-3}$) & \nodata & \nodata & 
1150$^{+2000}_{-1000} $ & 1200$\pm$550 \\
$T_e$([\ion{O}{3}]) (K) & 8150$\pm$150 & 8600$\pm$250 & 
14100$\pm$300 & 15700$\pm$300 \\
$T_e$([\ion{N}{2}]) (K) & 8600$\pm$450 & 8850$\pm$400 & 
12000$\pm$900 & 16500$\pm$1400 \\
\enddata
\end{deluxetable}

\clearpage

\begin{deluxetable}{lcccc}
\tablewidth{0pt}
\tablenum{4}
\tablecaption{Ionic abundances from collisionally excited lines\tablenotemark{a} \label{tbl-4}}
\tablehead{
\colhead{Ion} & \colhead{NGC 604} & \colhead{NGC 5461} & 
\colhead{NGC 5471} & \colhead{NGC 2363} }
\startdata
O$^{++}$/H$^+$ & 8.27$\pm$0.05 & 8.34$\pm$0.06 & 7.91$\pm$0.04 & 
7.85$\pm$0.04 \\
N$^+$/H$^+$ & 6.91$\pm$0.08 & 6.94$\pm$0.07 & 5.88$\pm$0.10& 
4.98$\pm$0.05 \\
S$^+$/H$^+$ & 5.90$\pm$0.08 & 5.82$\pm$0.08 & 5.37$\pm$0.09 & 
4.65$\pm$0.05 \\
S$^{++}$/H$^+$ & 6.91$\pm$0.05 & 6.75$\pm$0.09 & 5.97$\pm$0.07 & 
5.72$\pm$0.05 \\
Ar$^{++}$/H$^+$ & 6.20$\pm$0.05 & 6.20$\pm$0.05 & 5.45$\pm$0.03 & 
5.25$\pm$0.04 \\
Ar$^{3+}$/H$^+$ & 4.17$\pm$0.12 & 4.13: & 4.83$\pm$0.06 & 
5.15$\pm$0.05 \\
K$^{3+}$/H$^+$ & \nodata & \nodata & 3.86: & 4.14$\pm$0.10 \\
Fe$^{++}$/H$^+$ & 5.43$\pm$0.09 & 5.66$\pm$0.08 & 5.26$\pm$0.09 & 4.20$\pm$0.08 \\
Fe$^{3+}$/H$^+$ & \nodata & \nodata & 5.63: & 5.10: \\
\enddata
\tablenotetext{a}{In units of 12+log(X$^m$/H$^+$).}
\end{deluxetable}

\clearpage
\begin{deluxetable}{lcccc}
\tablewidth{0pt}
\tablenum{5}
\tablecaption{Ionic abundances from recombination lines\tablenotemark{a} \label{tbl-5}}
\tablehead{
\colhead{Ion} & \colhead{NGC 604} & \colhead{NGC 5461} & 
\colhead{NGC 5471} & \colhead{NGC 2363} }
\startdata
C$^{++}$/H$^+$ & 8.28$\pm$0.06 & 8.24$\pm$0.10 & 7.70:& 7.60: \\
O$^{++}$/H$^+$ & 8.47$\pm$0.10 & 8.63$\pm$0.12 & 8.12: & 
8.19$\pm$0.11 \\
\enddata
\tablenotetext{a}{In units of 12+log(X$^m$/H$^+$).}
\end{deluxetable}

\clearpage

\begin{deluxetable}{lcccccc}
\tabletypesize{\scriptsize}
\tablewidth{0pt}
\tablenum{6}
\tablecaption{Comparison of ionic abundances and $t^2$ parameter \label{tbl-6}}
\tablehead{
\colhead{} & \multicolumn{3}{c}{O$^{++}$/H$^+$} & \multicolumn{3}{c}{C$^{++}$/H$^+$} \\
\colhead{Object} & \colhead{CELs} & \colhead{RLs} & \colhead{$t^2$} &  
\colhead{CELs} & \colhead{RLs} & \colhead{$t^2$} 
}
\startdata
NGC 604 &   8.27$\pm$0.05 & 8.47$\pm$0.10 & 0.027$\pm$0.018 & 
\nodata & 8.28$\pm$0.06 & \nodata \\
NGC 5461 &  8.34$\pm$0.06 & 8.63$\pm$0.12 & 0.041$\pm$0.021 & 
8.27\tablenotemark{a} /7.90\tablenotemark{b} & 8.24$\pm$0.10 & 
$-$0.003/0.030 \\
NGC 5471 &  7.91$\pm$0.04 & 8.12: & 0.074: & 7.30\tablenotemark{a} /7.28\tablenotemark{b} & 7.70: & 0.091:/0.095: \\
NGC 2363 &  7.85$\pm$0.04 & 8.19$\pm$0.11 & 0.128$\pm$0.045 & 7.29\tablenotemark{c} & 7.60: & 0.102: \\
\multicolumn{7}{c}{Galactic \ion{H}{2} regions}\\
M~42 - 1\tablenotemark{d} & 8.39 & 8.52 & 0.018 & 8.05 & 8.36 & 0.025 \\
M~42 - 2\tablenotemark{d} & 8.39 & 8.59 & 0.028 & 7.94 & 8.34 & 0.031 \\
M~8\tablenotemark{e} & 8.02 & 8.32 & 0.040 & 7.95 & 8.33 & 0.028 \\
M~17 - 3\tablenotemark{f} & 8.49& 8.75 & 0.033 & \nodata & 8.57 & \nodata \\
M~17 - 14\tablenotemark{f} & 8.40 & 8.74 & 0.044 & \nodata & 8.69 & \nodata \\
\enddata
\tablenotetext{a}{Garnett et al. 1999 assuming $R_V$ = 3.1.}
\tablenotetext{b}{Garnett et al. 1999 assuming $R_V$ = 5.}
\tablenotetext{c}{Peimbert et al. 1986.}
\tablenotetext{d}{Esteban et al. 1998.}
\tablenotetext{e}{Esteban et al. 1999a.}
\tablenotetext{f}{Esteban et al. 1999b.}
\end{deluxetable}

\clearpage

\begin{deluxetable}{lcccc}
\tablewidth{0pt}
\tablenum{7}
\tablecaption{$N({\rm He})/N({\rm H})$\tablenotemark{a} \
for constant temperature \label{tbl-7}}
\tablehead{
\colhead{Ratio/Parameter} & \colhead{NGC 604} & \colhead{NGC 5461} & 
\colhead{NGC 5471} & \colhead{NGC 2363} }
\startdata
He$^+$/H$^+$ (4387 \mbox{\AA})& 965$\pm$97 &  904$\pm$90 & \
876$\pm$88 & 875$\pm$88 \\
He$^+$/H$^+$ (4471 \mbox{\AA})& 861$\pm$34 &  875$\pm$35 & \
854$\pm$34 & 811$\pm$32 \\
He$^+$/H$^+$ (4921 \mbox{\AA})& 909$\pm$45 &  913$\pm$46 & \
794$\pm$40 & 832$\pm$33 \\
He$^+$/H$^+$ (5876 \mbox{\AA})& 919$\pm$18 &  884$\pm$27 & \
913$\pm$27 & 903\tablenotemark{b} \\
He$^+$/H$^+$ (6678 \mbox{\AA})& 910$\pm$36 &  896$\pm$36 & \
819$\pm$33 & 803$\pm$32 \\
He$^+$/H$^+$ (7065 \mbox{\AA})\tablenotemark{c}& 904 &  880 & 
824 & 814 \\
$<$He$^+$/H$^+$$>$ & 905$\pm$18 &  889$\pm$17 & 
862$\pm$17 & 818$\pm$18 \\
He$^{++}$/H$^+$ & \nodata & 1.1$\pm$0.4 & 4.9$\pm$0.7 & 
3.6$\pm$0.4 \\
He$^{0}$/H$^+$& 36$\pm$17 & 52$\pm$18 & \nodata & 
\nodata\\
He/H & 941$\pm$25 &  942$\pm$25 & \
861$\pm$19 & 816$\pm$20 \\
\cutinhead{Adopted parameters}
$t^2$ & 0.00 & 0.00 & 0.00 & 0.00\\
$\tau(3889)$ & 1.0 & 5.0 & 0.2 & 1.5\\
$N_e$(He~{\sc{ii}}) (cm$^{-3}$)& 100 & 300 & 100 & 400\\
$ICF$(He) & 1.042 & 1.058 & 0.993 & 0.993\\
\enddata
\tablenotetext{a}{Given in units of $10^{-4}$, for the case of
no hydrogen collisions.}
\tablenotetext{b}{This value is not used to derive $<$He$^+$/H$^+$$>$, see text.}
\tablenotetext{c}{This line strongly depends on $\tau(3889)$ and $N_e$(He~{\sc{ii}}),
therefore it is not used to derive $<$He$^+$/H$^+$$>$, see text.}
\end{deluxetable}

\clearpage

\begin{deluxetable}{lcccc}
\tablewidth{0pt}
\tablenum{8}
\tablecaption{$N({\rm He})/N({\rm H})$\tablenotemark{a} \
for temperature variations \label{tbl-8}}
\tablehead{
\colhead{Ratio/Parameter} & \colhead{NGC 604} & \colhead{NGC 5461} & 
\colhead{NGC 5471} & \colhead{NGC 2363} }
\startdata
He$^+$/H$^+$ (4387 \mbox{\AA})& 955$\pm$95 &  891$\pm$89 & \
856$\pm$86 & 848$\pm$85 \\
He$^+$/H$^+$ (4471 \mbox{\AA})& 847$\pm$40 &  857$\pm$34 & \
835$\pm$33 & 792$\pm$32 \\
He$^+$/H$^+$ (4921 \mbox{\AA})& 892$\pm$45 &  890$\pm$45 & \
769$\pm$38 & 796$\pm$32 \\
He$^+$/H$^+$ (5876 \mbox{\AA})& 885$\pm$27 &  838$\pm$25 & \
867$\pm$26 & 849\tablenotemark{b} \\
He$^+$/H$^+$ (6678 \mbox{\AA})& 875$\pm$36 &  849$\pm$34 & \ 
771$\pm$31 & 737$\pm$29 \\
He$^+$/H$^+$ (7065 \mbox{\AA})\tablenotemark{c}& 894 &  845 & 
833 & 799 \\
$<$He$^+$/H$^+$$>$ & 878$\pm$17 &  854$\pm$17 & 
825$\pm$16 & 779$\pm$18 \\
He$^{++}$/H$^+$ & \nodata & 1.1$\pm$0.4 & 4.9$\pm$0.7 & 
3.6$\pm$0.4 \\
He$^{0}$/H$^+$& 37$\pm$17 & 50$\pm$17 & \nodata & 
\nodata\\
He/H & 915$\pm$24 &  905$\pm$24 & \
824$\pm$18 & 777$\pm$20 \\
\cutinhead{Adopted parameters} 
$t^2$ & 0.027 & 0.041 & 0.074 & 0.128\\
$\tau(3889)$ & 1.5 & 6.5 & 1.2 & 3.5\\
$N_e$(He~{\sc{ii}}) (cm$^{-3}$)& 200 & 300 & 100 & 600\\
$ICF$(He) & 1.042 & 1.058 & 0.993 & 0.993\\
\enddata
\tablenotetext{a}{Given in units of $10^{-4}$, for the case of
no hydrogen collisions.}
\tablenotetext{b}{This value is not used to derive $<$He$^+$/H$^+$$>$, see text.}
\tablenotetext{c}{This line strongly depends on $\tau(3889)$ and $N_e$(He~{\sc{ii}}),
therefore it is not used to derive $<$He$^+$/H$^+$$>$, see text.}
\end{deluxetable}

\clearpage

\begin{deluxetable}{lcccccccc}
\tabletypesize{\footnotesize}
\tablewidth{0pt}
\tablenum{9}
\tablecaption{Ionic and total gaseous abundances \label{tbl-9}}
\tablehead{
\colhead{Ion /} & \multicolumn{2}{c}{NGC~604} & 
\multicolumn{2}{c}{NGC 5461} & \multicolumn{2}{c}{NGC 5471} & \multicolumn{2}{c}{NGC 2363} \\
\colhead{Element} & \colhead{$t^2$ = 0.00} & \colhead{$t^2$ = 0.027} & \colhead{$t^2$ = 0.00} & \colhead{$t^2$ = 0.041} & \colhead{$t^2$ = 0.00} & \colhead{$t^2$ = 0.074} & \colhead{$t^2$ = 0.00} & \colhead{$t^2$ = 0.128} 
}
\startdata
C$^{++}$ & \nodata & 8.28$\pm$0.06 & \nodata & 8.24$\pm$0.10 & \nodata & 7.70: & 
 \nodata & 7.60: \\
N$^+$ & 6.91$\pm$0.08 & 7.01$\pm$0.08 & 6.94$\pm$0.07 & 7.10$\pm$0.07 & 
5.88$\pm$0.10 & 6.05$\pm$0.10 & 4.98$\pm$0.05 & 5.15$\pm$0.05 \\
O$^+$ & 8.09$\pm$0.11\tablenotemark{a} & 8.21$\pm$0.11\tablenotemark{a} & 8.17$\pm$0.08\tablenotemark{b} & 
8.34$\pm$0.08\tablenotemark{b} & 7.40$\pm$0.10\tablenotemark{b} & 7.58$\pm$0.10\tablenotemark{b} & 6.47$\pm$0.10\tablenotemark{c} & 6.65$\pm$0.10\tablenotemark{c} \\
O$^{++}$ & 8.27$\pm$0.05 & 8.47$\pm$0.10 & 8.34$\pm$0.06 & 
8.63$\pm$0.12 & 7.91$\pm$0.04 & 8.12: & 7.85$\pm$0.04 & 
8.19$\pm$0.11 \\
S$^+$ & 5.90$\pm$0.08 & 6.00$\pm$0.08 & 5.82$\pm$0.08 & 5.98$\pm$0.08 & 5.37$\pm$0.09 & 5.54$\pm$0.09 & 4.65$\pm$0.05 & 4.82$\pm$0.05 \\
S$^{++}$ & 6.91$\pm$0.05 & 7.13$\pm$0.05 & 6.75$\pm$0.09 & 7.08$\pm$0.09 & 5.97$\pm$0.07 & 6.20$\pm$0.07 & 5.72$\pm$0.05 & 6.08$\pm$0.05 \\
Ar$^{++}$ & 6.20$\pm$0.05 & 6.37$\pm$0.05 & 6.20$\pm$0.05 & 6.44$\pm$0.05 & 5.45$\pm$0.03 & 5.63$\pm$0.03 & 5.25$\pm$0.04 & 5.54$\pm$0.04 \\
Ar$^{3+}$ & 4.17$\pm$0.12 & 4.37$\pm$0.12 & 4.13: & 4.43: & 4.83$\pm$0.06 & 5.05$\pm$0.06 & 5.15$\pm$0.05 & 5.49$\pm$0.05 \\
K$^{3+}$ & \nodata & \nodata & \nodata & \nodata & 3.86: & 4.05: & 
4.14$\pm$0.10 & 4.45$\pm$0.10 \\
Fe$^{++}$ & 5.43$\pm$0.09 & 5.64$\pm$0.09 & 5.66$\pm$0.08 & 5.96$\pm$0.08 & 5.26$\pm$0.09 & 5.48$\pm$0.09 & 4.20$\pm$0.08 & 4.55 $\pm$0.08 \\
Fe$^{3+}$ & \nodata & \nodata & \nodata & \nodata & 5.63: & 5.84: & 5.10:& 
5.44: \\
\\
C & \nodata & 8.44$\pm$0.06 &  \nodata &  8.40$\pm$0.10 & \nodata &  7.79: & \nodata & 
7.62: \\
N & 7.31$\pm$0.14& 7.46$\pm$0.14 & 7.33$\pm$0.12 & 7.57$\pm$0.12 & 6.51$\pm$0.15 & 6.70$\pm$0.15 & 6.38$\pm$0.12 & 6.70$\pm$0.12 \\
O & 8.49$\pm$0.06 & 8.66$\pm$0.10 & 8.56$\pm$0.07 & 8.81$\pm$0.12 & 8.03$\pm$0.05 & 8.23: & 7.87$\pm$0.04 & 8.20$\pm$0.11 \\
S & 7.13$\pm$0.05 & 7.34$\pm$0.05 & 6.96$\pm$0.09 & 7.27$\pm$0.09 & 6.30$\pm$0.07 & 6.52$\pm$0.07 & 6.47$\pm$0.05 & 6.82$\pm$0.05 \\
Ar & 6.26$\pm$0.05 & 6.43$\pm$0.05 & 6.23$\pm$0.05 & 6.47$\pm$0.05 & 5.56$\pm$0.03 & 5.74$\pm$0.03 & 5.50$\pm$0.05 & 5.82$\pm$0.05 \\
Fe\tablenotemark{d} & 6.22$\pm$0.09& 6.43$\pm$0.09 & 6.45$\pm$0.08 & 6.75$\pm$0.08 & 6.14$\pm$0.09 & 6.36$\pm$0.09 & 5.81$\pm$0.08 & 6.16$\pm$0.08 \\
Fe\tablenotemark{e} &  \nodata & \nodata & \nodata & \nodata & 5.78: & 
5.99:& 5.16: & 5.49: \\
\enddata
\tablenotetext{a}{Line intensities taken from D\'\i az et al. 1987.}
\tablenotetext{b}{Line intensities taken from Torres-Peimbert et al. 1989.}
\tablenotetext{c}{Line intensities taken from Gonz\'alez-Delgado et al. 
1994.}
\tablenotetext{d}{Determined from Fe$^{++}$/H$^+$ and assuming an $ICF$(Fe$^{++}$)}
\tablenotetext{e}{Lower limits determined from Fe$^{++}$/H$^+$ + Fe$^{3+}$/H$^+$}
\end{deluxetable}

\clearpage

\begin{deluxetable}{lccc}
\tablewidth{0pt}
\tablenum{10}
\tablecaption{Abundance gradients in M101 and the Milky Way.
 \label{tbl-10}}
\tablehead{
\colhead{} & \colhead{O/H} & \colhead{C/H} & \colhead{C/O} \\
\colhead{Galaxy} & \multicolumn{3}{c}{(dex kpc$^{-1}$)}  
}
\startdata
M101\tablenotemark{a} &  $-$0.038 & $-$0.040: &  $-$0.002: \\
M101\tablenotemark{b} &  $-$0.030 & $-$0.055 &  $-$0.025  \\
M101\tablenotemark{c} &  $-$0.030 & $-$0.075 &  $-$0.040  \\
Milky Way\tablenotemark{d} & $-$0.049 & $-$0.086 &  $-$0.037 \\
\enddata
\tablenotetext{a}{This paper.}
\tablenotetext{b}{Garnett et al. 1999 ($R_V = 5.0$).}
\tablenotetext{c}{Garnett et al. 1999 ($R_V = 3.1$).}
\tablenotetext{d}{Esteban et al. 1999b, with revised 12 $+$ log (C/H) values
for Orion (8.43), M8 (8.58), and M17 (8.66); see text.}
\end{deluxetable}

\clearpage

\begin{deluxetable}{lcccccc}
\tabletypesize{\footnotesize}
\tablewidth{0pt}
\tablenum{11}
\tablecaption{ Abundances of Helium, Oxygen and Heavy Elements by Mass \label{tbl-11}}
\tablehead{
\colhead{} & \multicolumn{2}{c}{$Y$} & \multicolumn{2}{c}{$O$} & \multicolumn{2}{c}{$Z$}\\
\cline{2-3}\cline{4-5}\cline{6-7}
\colhead{Object} & \colhead{$t^2 = 0.00$} & \colhead{$t^2 > 0.00$} & \colhead{$t^2 = 0.00$} &  
\colhead{$t^2 > 0.00$} & \colhead{$t^2 = 0.00$} & \colhead{$t^2 > 0.00$} 
}
\startdata
NGC 604                   & 0.2705 & 0.2641 & 0.00427 & 0.00634 & 0.0108 & 0.0142 \\
NGC 5461                  & 0.2706 & 0.2612 & 0.00501 & 0.00896 & 0.0115 & 0.0175 \\
M17\tablenotemark{a}      & 0.2766 & 0.2677 & 0.00440 & 0.00849 & 0.0135 & 0.0201 \\
\enddata
\tablenotetext{a}{Peimbert \& Peimbert 2002.}
\end{deluxetable}

\clearpage

\begin{deluxetable}{lcccc}
\tablewidth{0pt}
\tablenum{12}
\tablecaption{Values of $\Delta Y/\Delta O$ and $\Delta Y/\Delta Z$
\label{tbl-12}}
\tablehead{
\colhead{} & \multicolumn{2}{c}{$\Delta Y$/$\Delta O$} & 
\multicolumn{2}{c}{$\Delta Y$/$\Delta Z$} \\
\cline{2-3}\cline{4-5}
\colhead{Object} & \colhead{$t^2 = 0.00$\tablenotemark{a}} & \colhead{$t^2 > 0.00$\tablenotemark{b}} 
& \colhead{$t^2 = 0.00$\tablenotemark{a}} & \colhead{$t^2 > 0.00$\tablenotemark{b}} 
}
\startdata
NGC 604              & $5.40\pm2.10$ & $4.05\pm1.50$ & $2.15\pm0.85$ & $1.80\pm0.65$ \\
NGC 5461             & $4.60\pm1.95$ & $2.55\pm1.05$ & $2.00\pm0.85$ & $1.30\pm0.55$ \\
M17\tablenotemark{c} & $6.60\pm1.25$ & $3.45\pm0.65$ & $2.15\pm0.45$ & $1.45\pm0.30$ \\
\enddata
\tablenotetext{a}{$Y_p$(+Hc)=0.2475, Peimbert et al. 2002}
\tablenotetext{b}{$Y_p$(+Hc)=0.2384, Peimbert et al. 2002}
\tablenotetext{c}{Peimbert \& Peimbert 2002.}
\end{deluxetable}

\clearpage

\begin{figure}
\plotone{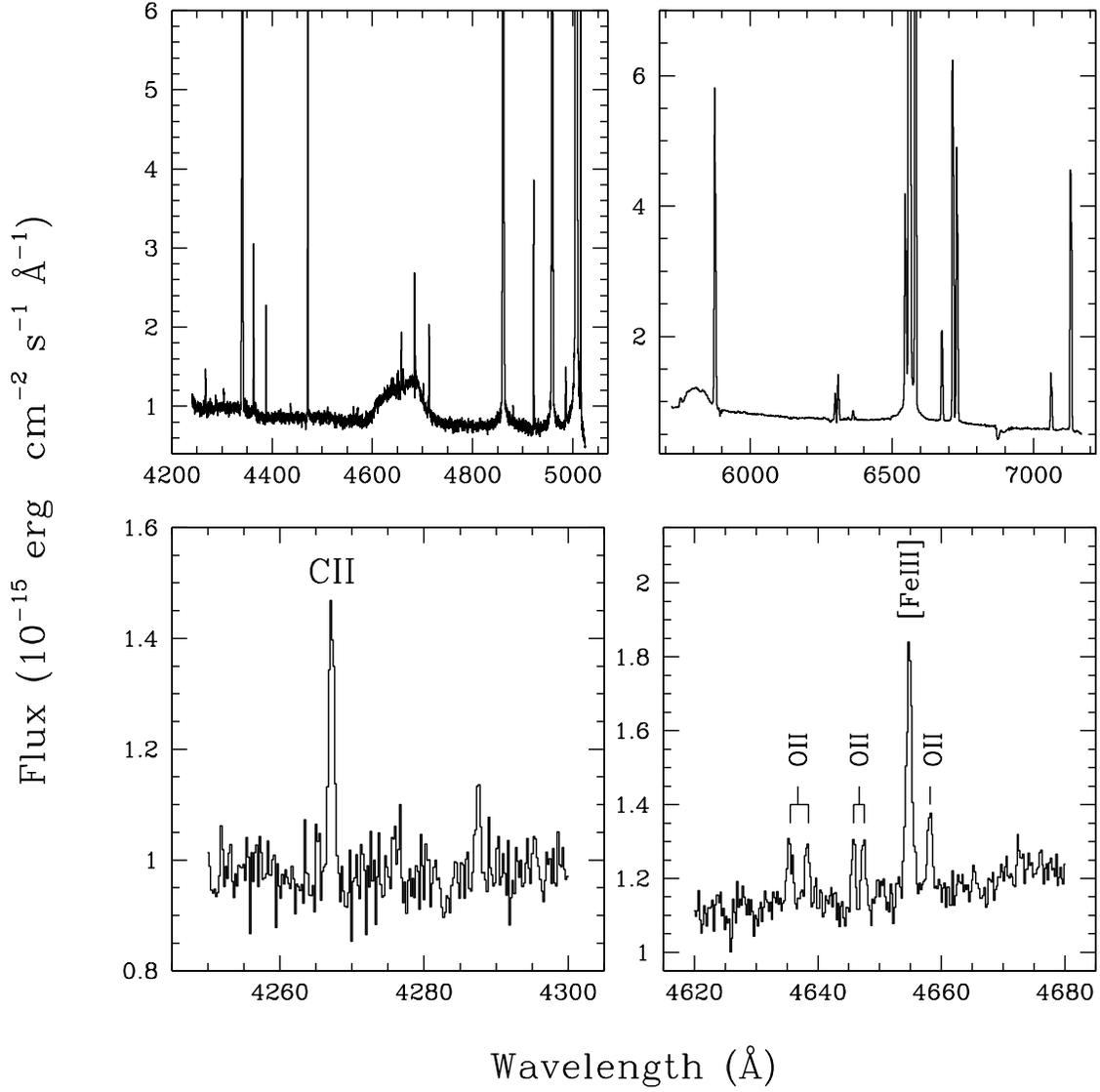}
\caption{Spectrum of NGC 604. The two upper panels show the whole spectral range observed expanded along the flux axis. The two lower panels 
show enlargements of the spectral zones containing \ion{C}{2} 4267 \mbox{\AA}\ line and the lines of multiplet 1 of \ion{O}{2} around 4650 \mbox{\AA}. \label{fig1}}
\end{figure}

\clearpage

\begin{figure}
\plotone{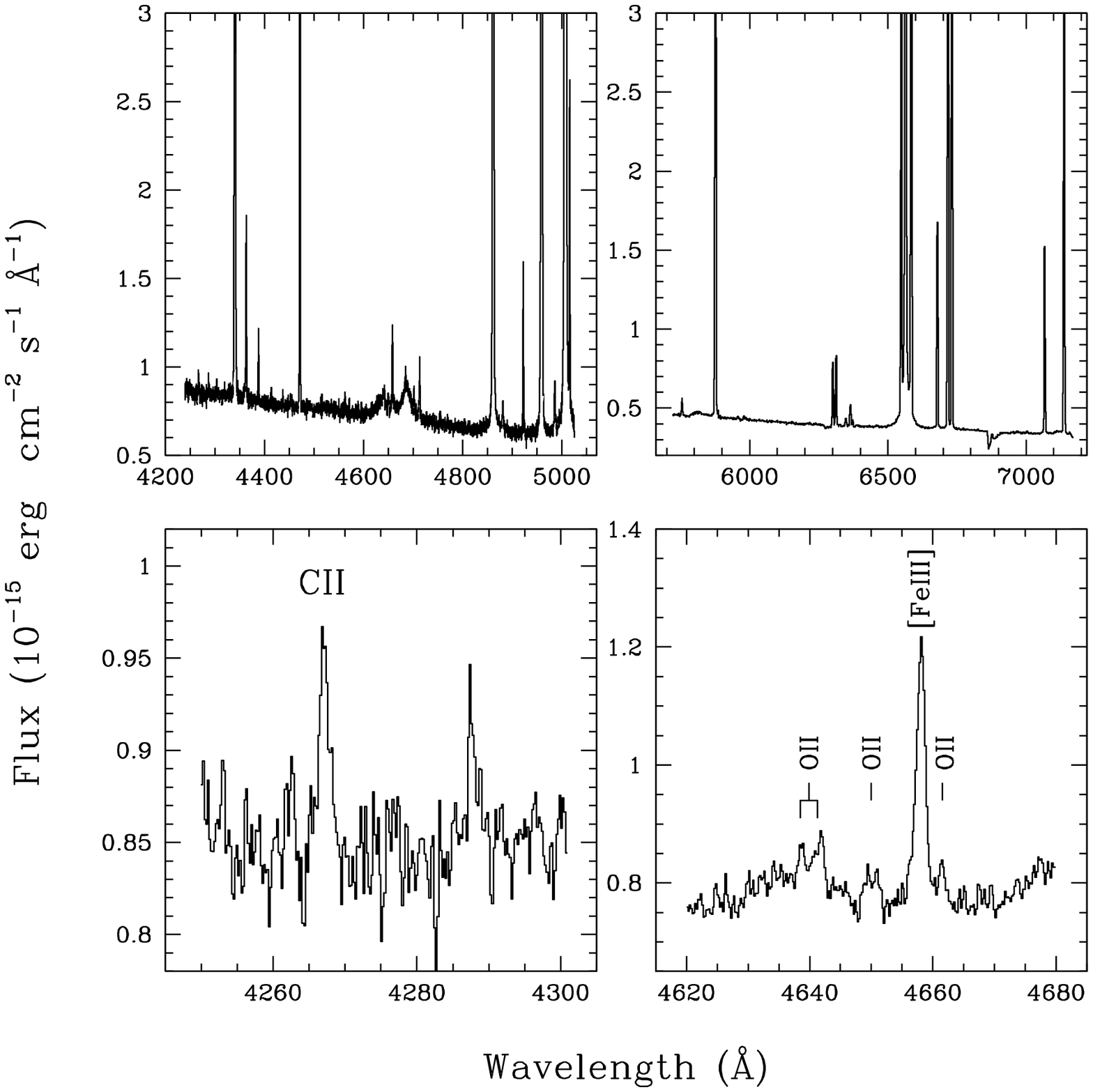}
\caption{The same as Fig. 1 but for NGC 5461\label{fig2}}
\end{figure}

\clearpage

\begin{figure}
\plotone{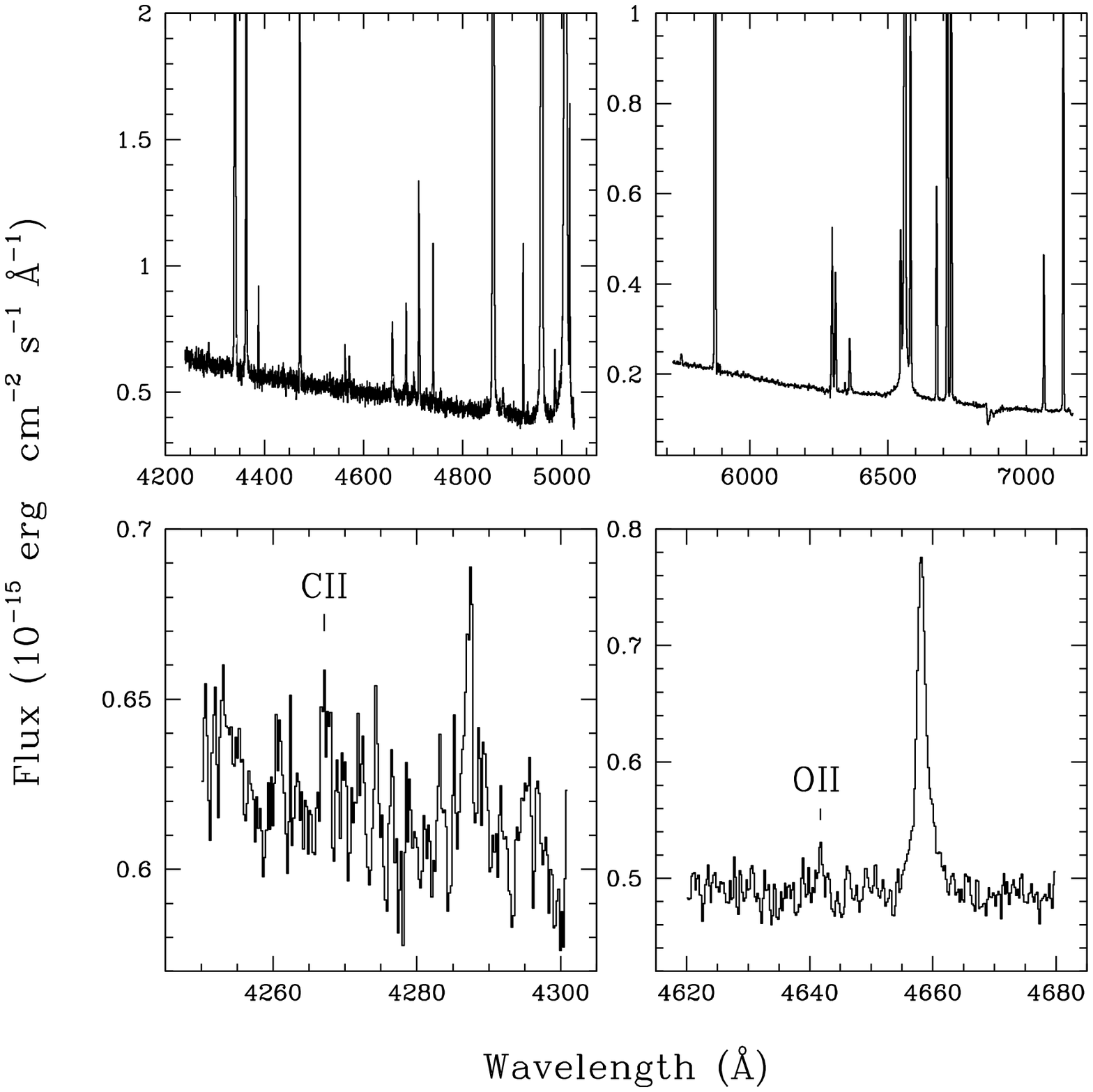}
\caption{The same as Fig. 1 but for NGC 5471 \label{fig3}}
\end{figure}

\clearpage

\begin{figure}
\plotone{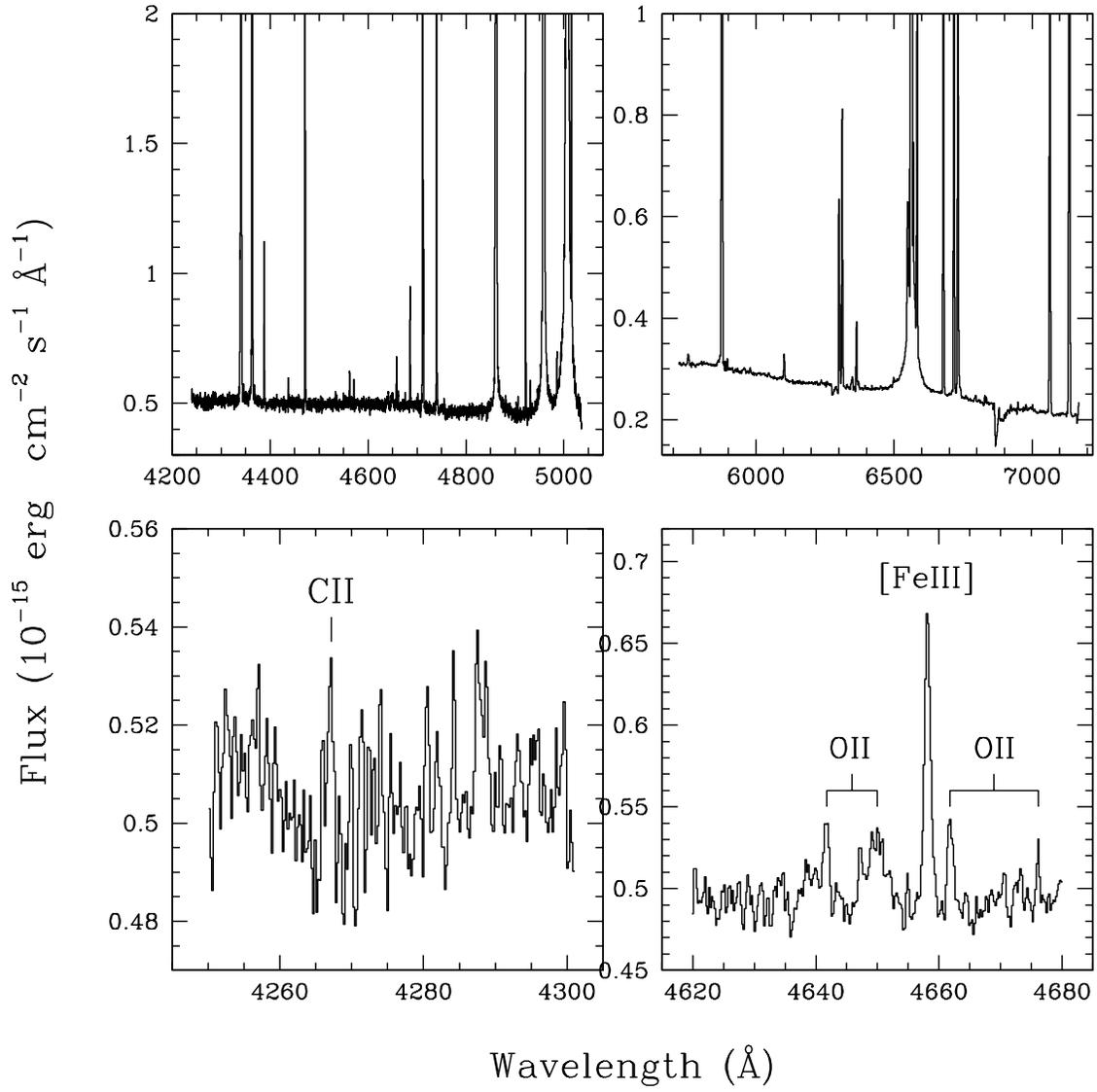}
\caption{The same as Fig. 1 but for NGC 2363 \label{fig4}}
\end{figure}

\clearpage

\begin{figure}
\plotone{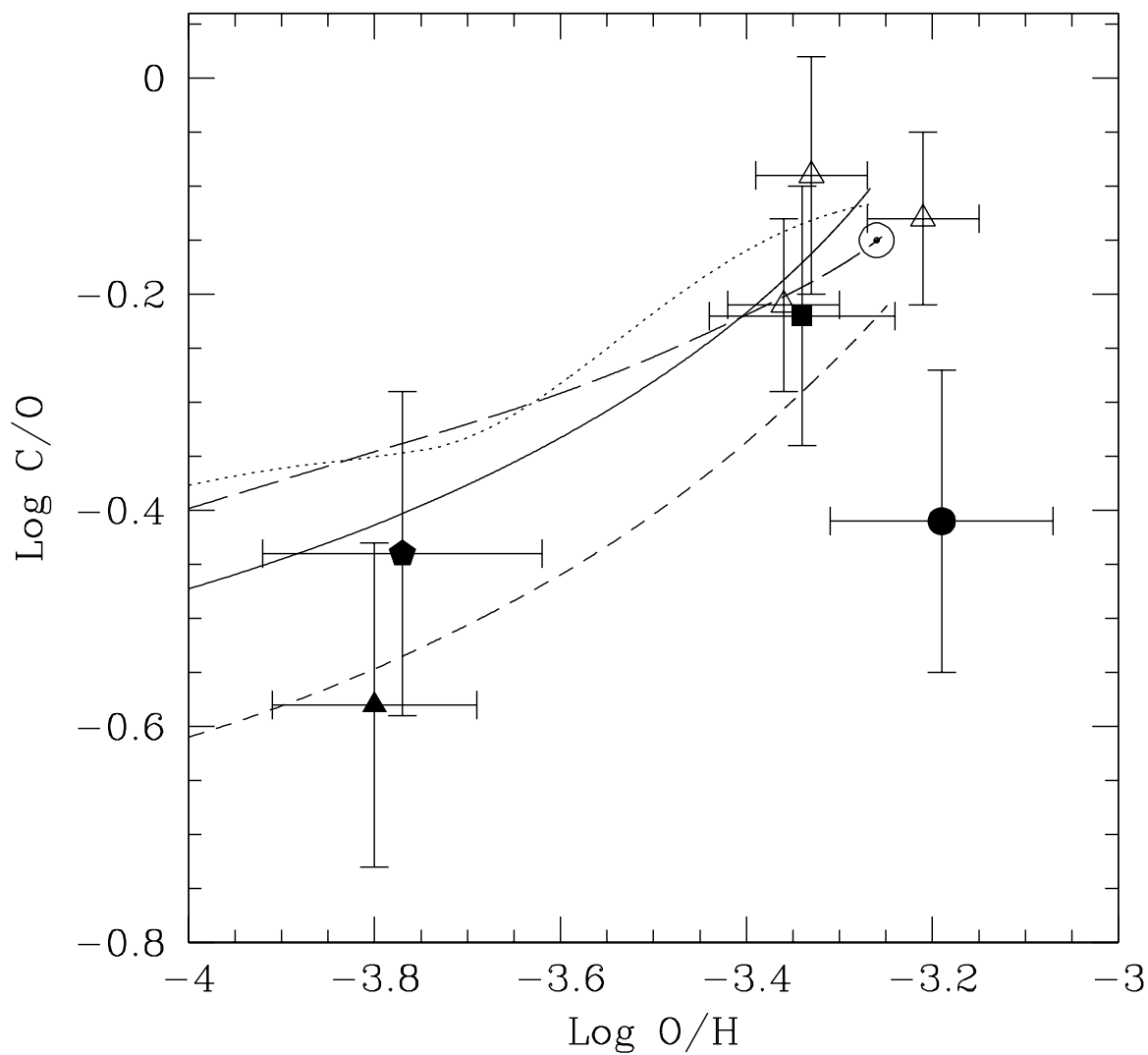}
\caption{C/O vs. O/H gaseous abundances for the objects included in this paper (full symbols; 
NGC~604: square, NGC~5461: circle, NGC~5471: pentagon, NGC~2363: triangle) and Galactic \ion{H}{2} regions (empty triangles: Orion nebula, 
M8 and M17). A nominal 40\% uncertainty have been assumed for C/O in 
the cases of  NGC~5471 and NGC~2363 as well as for O/H in the case of 
NGC~5471. The solar symbol represents the abundances of the Sun \citep{all01,hol01}. The lines show the predictions by chemical evolution 
models for the Galactic disk by \citet{car00}, see text for details. \label{fig5}}
\end{figure}

\end{document}